\documentclass[a4paper,fleqn,usenatbib]{mnras}

\usepackage{amsmath}	
\usepackage{amssymb}	

\usepackage{mathptmx}
\usepackage{txfonts}

\usepackage[T1]{fontenc}
\usepackage{ae,aecompl}
\usepackage{float}

\usepackage{times}

\usepackage{graphicx}
\usepackage{color}
\usepackage{bm}

\newcommand*{\sgra}{Sgr~A$^\star$}

\title[An MHD model for \sgra\ flares]{A magnetohydrodynamic model for multi-wavelength flares from Sagittarius~A$^\star$ (I): model and the near-infrared and X-ray flares}

\author[Li, Yuan, \& Wang]{
Ya-Ping Li$^{1,2}$\thanks{E-mail: leeyp2009@gmail.com (YPL)},
Feng Yuan$^{1}$\thanks{E-mail: fyuan@shao.ac.cn (FY)},
\& Q. Daniel Wang$^{3}$\thanks{E-mail: wqd@astro.umass.edu (QDW)}\\
$^{1}$Key Laboratory for Research in Galaxies and Cosmology, Shanghai Astronomical Observatory, Chinese Academy of Sciences, 80\\
Nandan Road, Shanghai 200030, China\\
$^{2}$Department of Astronomy and Institute of Theoretical Physics and Astrophysics, Xiamen University, Xiamen, Fujian 361005, China\\
$^{3}$Department of Astronomy, University of Massachusetts, Amherst, MA 01003, USA
 \\
}

\date{Accepted xxx. Received xxx; in original form xxx}

\pubyear{2016}

\begin{document}
\label{firstpage}
\pagerange{\pageref{firstpage}--\pageref{lastpage}}
\maketitle

\begin{abstract}
Flares from the supermassive black hole in our Galaxy, Sagittarius~A$^\star$ (\sgra), are routinely observed over the last decade or so. Despite numerous observational and theoretical efforts, the nature of such flares still remains poorly understood, although a few phenomenological scenarios have been proposed. In this work,  we develop the \citet{Yuan09} scenario into a magnetohydrodynamic (MHD) model for \sgra\ flares. This model is analogous with the theory of solar flares and coronal mass ejection in solar physics. In the model, magnetic field loops emerge from the accretion flow onto \sgra\ and are twisted to form flux ropes because of shear and turbulence. The magnetic energy is also accumulated in this process until a threshold is reached. This then results in a catastrophic evolution of a flux rope with the help of magnetic reconnection in the current sheet. In this catastrophic process, the magnetic energy is partially converted into the energy of non-thermal electrons. We have quantitatively calculated the dynamical evolution of the  height, size, and velocity of the flux rope, as well as the magnetic field in the flare regions, and the energy distribution of relativistic electrons in this process. We further calculate the synchrotron radiation from these electrons and compare the obtained light curves with the observed ones.  We find that the model can reasonably explain the main observations of near-infrared (NIR) and X-ray flares including their light curves and spectra. It can also potentially explain the frequency-dependent time delay seen in radio flare light curves.
\end{abstract}
\begin{keywords}
  black hole physics---accretion, accretion discs---Galaxy: centre---magnetic reconnection---(magnetohydrodynamics) MHD---radiation mechanisms: non-thermal.%
\end{keywords}

\section{Introduction}
Various observations have confirmed beyond reasonable doubt that our Galaxy hosts a supermassive black hole (SMBH), Sagittarius~A$^\star$ (\sgra), with its mass of $M_{\bullet}\approx4\times10^{6}~M_{\odot}$ (where $M_{\odot}$ is the solar mass; see reviews by \citealt{Genzel10}).  Multi-wavelength observations of \sgra\  reveal that its bolometric luminosity is $L_{\rm bol}\sim 10^{-9}L_{{\rm Edd}}$ (where $L_{\rm Edd}$ is the Eddington luminosity), which is five orders of magnitude lower than that predicted by a standard thin disc accretion at the Bondi accretion rate \citep{Baganoff03}. A number of  theoretical efforts have been made accompanied by these observational progresses (see reviews by \citealt{Genzel10} and \citealt{Yuan14}). We now understand that an advection-dominated accretion flow (ADAF) scenario works for \sgra. In this model, the low luminosity is due to both the low radiative efficiency and the strong mass loss via wind of the ADAF (\citealt{Yuan03,Yuan12a,Yuan15,Narayan12,Li13,Wang13,Gu15,Roberts17}). \sgra\ is thus an excellent laboratory for studying the accretion and ejection physics in such radiatively inefficient accretion flows (RIAFs), which are ubiquitous in the nearby universe.

\sgra\ is usually in a quiescent state, and occasionally interrupted by rapid flares (on timescales $\sim1~$~hr), most significantly in X-ray \citep{Baganoff01} and near-infrared (NIR; \citealt{Genzel03,Ghez04}).
Many such flares have been observed in various wavebands, including those detected by \emph{Chandra} (\citealt{Baganoff01,Eckart04,Eckart06,Aharonian08,Eckart08,Marrone08,Yusef08,Eckart12,Nowak12,Neilsen13,Ponti15,Yuan16}), \emph{XMM-Newton} (\citealt{Goldwurm03,Porquet03,Porquet05,Belanger05,Yusef06a,Porquet08,Yusef09,Trap11,Ponti15}), \emph{Swift} (\citealt{Degenaar13,Degenaar15}),  and \emph{NuSTAR} (\citealt{Barriere14,Dibi14}).
Many have also been detected in NIR (\citealt{Genzel03,Eckart04,Eckart06,Ghez04,Yusef06a,Eckart08,Yusef08,Yusef09,Kunneriath10,Trap11,Eckart12,Haubois12,Hora14,Shahzamanian15}),
in sub-millimeter (\citealt{Eckart06,Yusef06a,Yusef09,Kunneriath10,Trap11,Eckart12,Haubois12,Dexter14,Bower15,Brinkerink15}),
and also in radio (\citealt{Yusef06b,Yusef08,Yusef09,Bower15,Brinkerink15}).

We now briefly summarize the main properties of these multi-wavelength flares (see also \citealt{DoddsEden09} for a detailed summary of the general properties of the NIR and X-ray flares from \sgra). The flare rate is roughly two per day in X-ray \citep{Neilsen13,Ponti15,Yuan16} and more frequently in NIR. The NIR and X-ray flares occur simultaneously within $3~\rm mins$ when both are observed in company (e.g., \citealt{Eckart04}, but see \citealt{Yusef12} for a counter-example). The amplitude of the NIR and X-ray flares can be up to $\sim 20$ and 160 of the quiescent fluxes, respectively \citep[e.g.,][]{Yusef09,Nowak12}.  The full width at half maximum (FWHM) of the NIR flare profile is about 60 mins, which is about twice that of the X-ray. There are substructure variations with characteristic timescale of $\sim20~\rm mins$ in the NIR light curve occasionally, but not present at the same level in X-ray \citep{DoddsEden09}. The light curves of both X-ray and NIR flares are roughly symmetric, but the brightest flare in \citet{Nowak12} shows remarkable asymmetry profile with a faster decline than rise. A large fraction of the X-ray flares in the XVP campaign also shows a faster rise and slow decay profile \citep{Yuan16}. The flares in the NIR are significantly polarized with the typical polarization degrees of the order of $20\pm10\%$ \citep[e.g.,][]{Eckart06a,Shahzamanian15}.

The X-ray radiation of \sgra\ is believed to have two distinctive states. One is a steady quiescent emission which is dominated by the radiation around the Bondi radius, while the other is point-like flare emission arisen from the innermost region of the accretion flow \citep[e.g.,][]{Baganoff01,Baganoff03,Wang13}. Therefore, the X-ray flares look like to be large amplitude, short duration events overlayed on a flat baseline \citep{Neilsen13,Li15,Yuan16}. However, it is less clear that such a conclusion can be applied to the NIR emission. Much of the NIR variation could just represent the red noise of the underlying quiescent emission \citep{Meyer08,Do09}.
Flares in sub-millimeter and radio show much shallower and  broader profiles than those in NIR and X-ray. There is a general trend that the peak flare emission at a higher radio frequency leads that of a lower one, e.g., 43 GHz leading 22 GHz by $20\sim40~{\rm mins}$ \citep{Yusef06b}.  There are also some evidence for time lags among radio, sub-millimeter and NIR/X-ray \citep{Marrone08,Yusef08,Yusef09,Brinkerink15}, although some debates on the correlation of variabilities at different bands exist \citep[e.g.,][]{DoddsEden10}.

Most theoretical works of the flares focus on their radiation mechanisms. The proposed flare models usually invoke synchrotron and/or inverse Compton radiation processes.
The highly polarized NIR emission is the evidence for a synchrotron origin of the NIR flares, produced by a population of non-thermal electrons (e.g., \citealt{Eckart06a,Shahzamanian15}, see also references therein). The non-thermal electrons are likely accelerated by magnetic reconnection, shock or turbulence in either an accretion flow (\citealt{Yuan04,DoddsEden09,Li15}, and references therein) or an assumed jet (\citealt{Markoff01}). The suggested radiation mechanisms for the X-ray flares includes synchrotron \citep{Markoff01,Yuan03,Yuan04,DoddsEden09}, inverse Compton scattering \citep{Markoff01,Yuan03,Eckart04,Eckart06,Liu06a,Marrone08,Yusef12}, and/or bremsstrahlung \citep{Liu02}. Continuous injections of a population of high-energy electrons are in general required  for the synchrotron mechanism in order to balance the fast cooling of the X-ray synchrotron emission. This may be a natural ingredient for the magnetohydrodynamic (MHD) process, e.g., magnetic reconnection considered in this work.


There do exist some theoretical models aimed to explain the physical origin of the flares, including accretion instabilities \citep{Tagger06,Falanga08}, orbiting hot spots \citep{Broderick05,Meyer06,Trippe07,Hamaus09}, expanding plasma blobs \citep{Yusef06b,Eckart06,Yusef09,DoddsEden10,Trap11}, and tidal disruption of asteroids by the SMBH \citep{Cadez08,Kostic09,Zubovas12}. Recently, modeling efforts have been focused on  magnetic reconnection within the accretion flow and are usually based on the MHD numerical simulations (\citealt{Chan09,Dexter09,Maitra09,DoddsEden10,Chan15,Ball16}). The flares are produced by the radiation of non-thermal electrons accelerated in the reconnection. Note that the ingredient of the invoked particle acceleration process in these modeling efforts (including the present one) have to be phenomenological in the sense that it is taken from other independent works, usually particle-in-cell simulations. One representative work is \citet{DoddsEden10}. In this work, they consider the synchrotron radiation from a population of non-thermal electrons transiently accelerated by an episodic magnetic reconnection occurred in the accretion flow. They assume time-dependent profiles for the injection rate and the magnetic field strength in the description of the non-thermal electron distribution evolution, which are responsible for the light curves of the NIR and X-ray flares, as well as the spectral energy distribution (SED). More recently, \citet{Ball16} find that X-ray variability could result from non-thermal electrons in localized highly magnetized regions, based on their general relativistic MHD (GRMHD) simulations. Most of these works suggest that magnetic reconnection likely plays an important role in producing the NIR, and especially X-ray flares\footnote{Some other possibilities have also been proposed, such as accretion instabilities \citep{Tagger06,Falanga08} and shocks \citep{Dexter13}.}.


In the present work, we propose an alternative model for the flares of \sgra. Different from the works mentioned above, we assume that the flares are caused by magnetic reconnection occurred not in the main body of the accretion flow, but instead in the surface or the coronal region of the accretion flow. This work is a development of the MHD model for the formation of episodic jets proposed by \citet{Yuan09}. The model is analogous to the coronal mass ejection (CME) of the Sun \citep{LF00}. Whereas the work by \citet{Yuan09} focuses on the dynamics of the ejection of blobs from the black hole accretion flow, the present work moves a step further to model the associated radiation. The modeling is motivated by the following facts: First, based on a statistical analysis for X-ray flares of \sgra, \citet{Li15} have shown that they are consistent with events from a three-dimensional self-organized criticality (SOC) system, similar to solar flares, which are powered by magnetic reconnection in the corona above the accretion flow.  Second, the detailed study to the solar flares over many years have shown that the powerful CMEs are physically associated with the strong solar flares, i.e., these flares and CMEs are likely the different manifestations of the same physical process. In fact, the flare model to be presented in this work is similar to the standard model of solar flares. Third, if the \sgra\ flares seen in NIR and X-ray are physically associated with those seen in radio, then we can naturally speculate that they are all linked to the plasmoid ejection process, as clearly indicated by the observed wavelength-dependent time lag mentioned above \citep[e.g.,][]{Yusef06b,Marrone08,Yusef08,Yusef09,Brinkerink15}. Lastly,  extensive MHD numerical simulations of black hole accretion flows have led to the consensus that the
hot accretion flow is enveloped by a tenuous corona which is magnetically dominated (see \citealt{Yuan14} for a review). The contrast of the density and magnetic-to-gas pressure ratio between the accretion flow and the corona is similar to that between the solar photosphere and solar corona \citep{Aschwanden05}. These facts provide us with the motivation to make the above analogy between the Sun and the accretion flow.

In fact, several works have been presented along the above line.  Without considering the dynamics of the ejection of a blob from the accretion flow, \citet{Kusunose11} calculated the synchrotron radiation from such a blob, assuming some time-dependent profiles for the non-thermal electron injection. The results were then compared with the observed light curve and SED of the NIR and X-ray flares in \sgra. More recently, \citet{Younsi15} investigated the emissions from the plasmoids with dynamics as described in the CME scenario for episodic jets. They considered special and general relativistic effects (especially the gravitational lensing one), on the light curves, but without paying attention to the spectra. Furthermore, \citet{Meng14} applied a similar CME model to interpreting the giant flare of three magnetars by assuming the free magnetic energy released in an eruptive process to power the giant flare events.

Here, we present a time-dependent MHD model for the flares of \sgra\ within the framework described by \citet{Yuan09}. We will calculate both the dynamics of the blob ejection and the corresponding radiation, including the light curves and spectrum of NIR and X-ray flares. In a subsequent work (Li et al. in preparation), we will focus on interpreting radio flares based on the same model. As we will see, the model can explain the observations quite well.
The paper is organized as follows. We describe our MHD model in Section~2. Numerical results of our model and comparisons with observations are presented in Section~3. We discuss our results in Section~4 and summarize the work in Section~5. Throughout this work, we assume that the SMBH mass is $M_{\bullet}=4\times10^6~M_{\odot}$ and its distance is $d=8~{\rm kpc}$.

\section{Methodology}

\subsection{Dynamical Evolutions}\label{sec:dyn}

\begin{figure}
\centering
\includegraphics[width=0.45\textwidth]{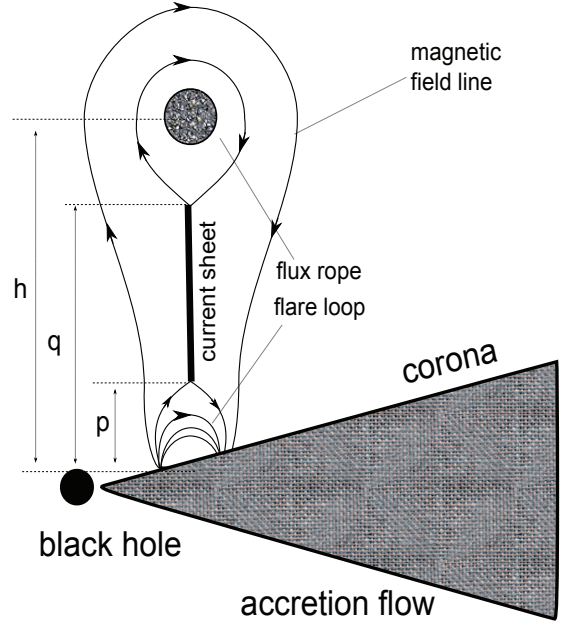}
\caption{Diagram of the magnetic field configuration, showing the mathematical notation used in the text. The solid lines with arrows present the magnetic field lines. The flux rope is denoted by the closed field line region enclosing a shaded circle (the core of the flux rope region). The semicircle region below the current sheet is named as the flare loop. The current sheet region is represented by the thick solid line.  The $x$-axis is parrel to the equatorial plane of the accretion flow, and the $y$-axis points upward. The bottom, top tip of the current sheet, and the height of the centre of the flux rope are denoted by $p$, $q$, and $h$, respectively. All these heights are measured from the disc surface in our calculations. The distance between the two magnetic source regions on the accretion flow is 2$\lambda$. }\label{fig:diagram}
\end{figure}

In this section we calculate the dynamical evolution of plasmoids ejected from the coronal region of the accretion flow. Our calculation starts with a newly formed plasmoid (also called flux rope), as shown by Figure~\ref{fig:diagram}, in which the closed shaded circle represents the core of the plasmoid. The formation of the plasmoid is not addressed here, but should be similar to the formation of the prominence in the Sun. It could be due to the thermal instability of the gas in the corona, or due to the reconnection of a magnetic loop emerged from the accretion flow into the corona (\citealt{Yuan09}).
The magnetic loops emerged from the accretion flow into the corona has been proposed as one of the main magnetic field configurations by \citet{Blandford02} and studied by some previous works (e.g., \citealt{Uzdensky08,Guan11})
\footnote{Using shearing box MHD numerical simulation approach, \citet{Guan11} find the existence of magnetic loops in the coronal region of the disc. Although this simulation is for a thin disc, we believe that the  presence of the magnetic loops should remain the same for a thick accretion flow. }.

Initially, the flux rope is enveloped with magnetic loops, with their foot points anchored in the accretion flow, as shown in Figure~\ref{fig:diagram}.
The subsequent dynamical evolution of the system due to the turbulent motion of the accretion flow consists of two stages connected with a triggering process. While the details can be found in \citet{Yuan09}, here we briefly summarize the whole process. In the first stage, the plasmoid is in an equilibrium when there is a balance among the forces due to the magnetic compression, magnetic tension, and gravity. The foot points of the magnetic field lines are anchored into the accretion flow. The magnetic energy is gradually accumulated in the coronal magnetic field in response to the turbulent motion and differential rotation of the flow. The evolution of the system in this stage is ideal, which means that magnetic reconnection does not take place in the corona. The gradual evolution of the boundary conditions in the accretion flow will inevitably bring the plasmiod into a critical point when the total magnetic energy of the system reaches a threshold, after which further evolution causes the loss of equilibrium of the system initiated by the triggering process. In the second stage, the loss of this equilibrium leads to a catastrophic evolution of the plasmoid, during which most of stored free magnetic field energy is rapidly released due to the magnetic reconnection. It is thus this non-ideal MHD process in this stage that powers the radiative flares and is our focus in the following calculations.

Figure~\ref{fig:diagram} shows an illustration diagram characterizing the disrupting process. The bolded vertical line represents the current sheet, a neutral region separating magnetic field lines of opposite polarity. The motion of the flux rope is mainly subject to the gravity and magnetic forces \citep{Lin06}. The gas pressure is assumed to be much smaller compared to the magnetic pressure in the coronal region and is thus neglected in this model.

After the loss of the equilibrium, the flux rope is thrust outward and its motion is governed by

\begin{eqnarray}
  m \gamma_{\rm b}^{3} \frac{d^{2}h}{d t^{2}} = \frac{1}{c}|\textbf{\emph{I}}
 \times\textbf{\emph{B}}_{\rm ext}|- F_{\rm g}
 \label{eq:Fmg}
\end{eqnarray}%
to the first order of approximation\footnote{This is in reference to the radius of the flux rope core $r_{00}$ being much less than the typical eruption length scale of the system $h$, i.e., $r_{00}/h\ll1$ \citep{Forbes91,Isenberg93}, which is always satisfied in our model.}, where \emph{m} is the total mass inside the flux rope per unit length, $\gamma_{_{\rm b}}=1/ \sqrt{1-\dot{h}^{2}/c^{2}}$ is its Lorentz factor, \emph{h} is the height of the flux rope from the surface of the accretion flow,
$\textbf{\emph{I}}$ is the total electric current intensity flowing inside the flux rope, $\textbf{\emph{B}}_{\rm ext}$ is the total external magnetic field measured at the centre of the flux rope. The first term on the righthand side of Equation~(\ref{eq:Fmg}) is the magnetic force and the second term $F_{\rm g}$ is the gravitational force acting on the mass inside the flux rope, both of which will be given below.


In the zeroth-order approximation, the following equations
hold \citep[see][for details]{LF00}:
\begin{eqnarray}
  \textbf{\emph{j}} \times \textbf{\emph{B}}=0,
  \label{eq:jbE}
\end{eqnarray}%
\begin{eqnarray}
  \textbf{\emph{j}}=\frac{c}{4\pi} \nabla \times \textbf{\emph{B}},
  \label{eq:dbE}
\end{eqnarray}%
where $\textbf{\emph{j}}$ and $\textbf{\emph{B}}$ are the electric current density and
the magnetic field in the system, respectively.

In the Cartesian coordinate system $(x, y)$, the $x$-axis is parallel to the equatorial plane of the accretion flow and $y$-axis points upward (see also Figure~\ref{fig:diagram}). Solving Equations (\ref{eq:jbE}) and (\ref{eq:dbE}) gives the configuration of the force-free magnetic field of the system \citep{Reeves05}
\begin{eqnarray}
  B(\zeta)=\frac{2 i A_{0}\lambda(h^{2}+\lambda^{2})\sqrt{(\zeta^{2}+p^{2})(\zeta^{2}+q^{2})} }
  {\pi (\zeta^{2}- \lambda^{2})(\zeta^{2}+h^{2})\sqrt{(\lambda^{2}+p^{2})(\lambda^{2}+q^{2})}},
  \label{eq:Bxy}
\end{eqnarray}%
where $\zeta=x+iy$, $A_{0}= B_{0}\pi \lambda_{0}$ is the source field strength and $B_{0}= 2 I_{0}/ ( c \lambda_{0} )$ is the normalization of the magnetic field strength on the surface of the accretion flow.

The corresponding vector potential function $\emph{A}(\zeta)$ is given by
\begin{eqnarray} \label{eq:Axy}
 A(\zeta)&=&  - \int {B(\zeta)} d\zeta .
\end{eqnarray}

Knowing the magnetic field configuration of the system, we can derive the forces acting on the flux rope. The magnetic force, $F_{\rm m}$, can be expressed as
\begin{eqnarray}\label{eq:Fm}
{F_{\rm m}} = \frac{{{B_0}^2{\lambda ^4}}}{8h{L^2_{\rm PQ}}}\left[\frac{{H^2_{\rm PQ}}}{{2{h^2}}} - \frac{({p^2}
 + {\lambda ^2})({h^2} - {q^2})}{{h^2} + {\lambda ^2}}\right. \nonumber \\
 -\left.\frac{({q^2} + {\lambda ^2})({h^2} - {p^2})}{{h^2} + {\lambda ^2}} \right],
\end{eqnarray}%
and the gravity, $F_{\rm g}$, is
\begin{eqnarray}\label{eq:Fg}
{F_{\rm g}} = \frac{{G{M_{\bullet}}\gamma_{\rm b}{m}}}{{{{\left( {{R_0} + h} \right)}^2}}},
\end{eqnarray}%
where $L_{\rm PQ}^{2}=(\lambda^{2}+p^{2})(\lambda^{2}+q^{2})$,
$H_{\rm PQ}^{2}=(h^{2}-p^{2})(h^{2}-q^{2})$, $\lambda$ is the half-distance between the two field line foot points anchored on the disc surface (see Figure~\ref{fig:diagram} for the illustration of these parameters), the initial enclosed mass per unit length is $m=m_{0}$  in the flux rope, and is given by $m_{0}=\xi\pi{r^2_{00}}n_{\rm e}m_{\rm H}$, $r_{00}$ the radius of the flux rope and usually $r_{00}=0.1\lambda$ is adopted according to the experience of the solar flares, $\xi$ the density ratio for the flux rope with respect to the background density $n_{\rm e}$, $m_{\rm H}$ is the mass of the Hydrogen atom, and $M_{\bullet}$ is the mass of the central black hole.

The first term in the square brackets on the righthand side of  Equation~(\ref{eq:Fm}) denotes the magnetic compression force, while the other two terms in the bracket represent the magnetic tension forces. It is this magnetic compression force that pushes the flux rope outwards (upwards), and makes the catastrophic loss of the equilibrium in the system possible. The magnetic tension and gravity force tends to pull the flux rope backwards (downwards). The system is in equilibrium initially as they balance each other. When the compression term
dominates over the other, the system loses its equilibrium, and thrusts the flux rope outward in a catastrophic fashion.

The dynamic equation governing the motion of the flux rope can thus be deduced to
\begin{eqnarray}\label{eq:dh2dt}
\gamma^{3}_{\rm b}\frac{d^{2}h}{d t^{2}}&=&\frac{{{B_0}^2{\lambda ^4}}}{8h{m}{L^2_{\rm PQ}}}
\left[\frac{{H^2_{\rm PQ}}}{{2{h^2}}} - \frac{({p^2} + {\lambda ^2})({h^2}- {q^2})}{{h^2} + {\lambda ^2}} \right. \nonumber \\
&-& \left. \frac{({q^2} + {\lambda ^2})({h^2} - {p^2})}{{h^2} + {\lambda ^2}}\right]
-\frac{G{M_{\bullet} }\gamma_{\rm b}}{{{\left( {{R_0} + h} \right)}^2}}.
\end{eqnarray}%

From Faraday's Law, the electric field in the reconnection region is induced in the reconnection process $E_{z}(t)$ and is given by
\begin{eqnarray} \label{eq:Ez}
E_{z}(t)= -\frac{1}{c}\frac{\partial A_{0}^{0}}{\partial t}
= M_{\rm A}V_{\rm A}B_{y}(0,y_{0})/c,
\end{eqnarray}%
where $A_{0}^{0} =A(0,p\leqslant y \leqslant q)$ is the magnitude of the vector potential along the current sheet, $y_{0}=(p+q)/2$ is the height of the current sheet centre, $V_{\rm A}\equiv{B}_{y}(0,y_{0})/\sqrt{4\pi\rho(y_{0})}$ is the local Alfv\'{e}n speed, $M_{\rm A}$ is the Alfv\'{e}n Mach number of the reconnection inflow and is a measure of the reconnection rate in the current sheet \citep{LF00}, which is defined to be the reconnection inflow speed $V_{\rm in}$ divided by the local Alfv\'{e}n speed $V_{\rm A}$ near the reconnection region (i.e., $M_{\rm A}\equiv V_{\rm in}/V_{\rm A}$). In this work, it is taken to be a constant measured at the current sheet centre $y_{0}$, and the magnetic field $B_{y}(0,y_{0})$ can be directly obtained from Equation~(\ref{eq:Bxy}) with $x=0$ and $y=y_{0}$. The expression of the density profile $\rho(y)$ will be given below.


As the system evolves dynamically and the ejector moves upward at speed $\dot{h}$, the electric field $E_{z}$ in Equation (\ref{eq:Ez}) can be written as
\begin{eqnarray}
E_{z}(t)&=&-\frac{1}{c}\frac{\partial A_{0}^{0}}{\partial t}\label{eq:dA00}
= -\frac{1}{c}\frac{\partial A_{0}^{0}}{\partial h}\dot{h}\nonumber\\
&=&-\frac{\dot{h}}{c}\left(\frac{\partial A_{0}^{0}}{\partial p}p'+ \frac{\partial A_{0}^{0}}
{\partial q}q'+ \frac{\partial A_{0}^{0}}{\partial h} \right)\nonumber\\
&=&-\frac{2I_{0}\dot{h}}{c^2}(A_{\rm 0p}p'+ A_{\rm 0q}q'+ A_{\rm 0h}),
\end{eqnarray}
where $\dot{h}= dh/dt$, $p'=dp/dh$ and $q'=dq/dh$, $A_{\rm 0p}$, $A_{\rm 0q}$, and $A_{\rm 0h}$ are given in the appendix.

In order to solve Equation~(\ref{eq:Fmg}), another equation is required, namely the frozen magnetic flux condition on the surface of the flux rope. The frozen-flux condition can be expressed as
\begin{eqnarray}
\frac{2 I_{0}}{c}A_{\rm R}= A(0, h- r_{0})= \textrm{const.}\label{eq:ffc},
\end{eqnarray}%
where $r_{0}$ is the radius of the flux rope. Taking the total derivative about $h$ on the both side of Equation (\ref{eq:ffc}) gives
\begin{eqnarray}
\frac{\partial A_{\rm R}}{\partial p}p'+ \frac{\partial A_{\rm R}}{\partial q}q'
+\frac{\partial A_{\rm R}}{\partial h}
= A_{\rm Rp}p'+ A_{\rm Rq}q'+ A_{\rm Rh} = 0. \nonumber \\ \label{eq:dAR}
\end{eqnarray}%
Then $p'$ and $q'$ can be obtained from (\ref{eq:dA00}) and (\ref{eq:dAR}):
\begin{eqnarray}
p'=\frac{\tilde{A_{\rm 0h}}A_{\rm Rq}- A_{\rm Rh}A_{\rm 0q}}{A_{\rm Rp}A_{\rm 0q}- A_{\rm 0p}A_{\rm Rq}},
\label{eq:dpdh}
\end{eqnarray}%
\begin{eqnarray}
q'=\frac{A_{\rm Rh}A_{\rm 0p}- \tilde{A_{\rm 0h}}A_{\rm Rp}}{A_{\rm Rp}A_{\rm 0q}- A_{\rm 0p}A_{\rm Rq}},
\label{eq:dqdh}
\end{eqnarray}%
where
\begin{eqnarray}\label{eq:A0hp}
\tilde{A_{\rm 0h}}= \frac{c E_{z}}{B_{0}\lambda \dot{h}}+ A_{\rm 0h}
= \frac{M_{\rm A}V_{\rm A}B_{y}(0,y_{0})}{B_{0}\lambda \dot{h}}+ A_{\rm 0h},
\end{eqnarray}
where most of these symbols have its nominal meaning. The other terms used in the above equations are shown in the appendix.
Then two equations governing the motions of the current sheet can be expressed as
\begin{eqnarray}
\frac{{dp}}{{dt}} = p' \dot{h}, \label{eq:dpdt} \\
\frac{{dq}}{{dt}} = q' \dot{h}. \label{eq:dqdt}
\end{eqnarray}%

\citet{Lin06} noted that a large amount of plasma in the corona is brought into the flux rope due to magnetic reconnection as the eruption evolves. The evolution of the total mass in the flux rope is governed by:
\begin{eqnarray}\label{eq:dmdt}
\frac{dm}{dt} &=& B_{0}M_{\rm A}\sqrt{\frac{n_{\rm e}m_{\rm H}}{\pi}}
\frac{\lambda^{2}(q-p)(h^{2}+\lambda^{2})}{(h^{2}-y_{0}^{2})
(y_{0}^{2}+\lambda^{2})} \nonumber \\
&\times& \sqrt{\frac{f(y_{0})(q^{2}-y_{0}^{2})(y_{0}^{2}-p^{2})}{(p^{2}+\lambda^{2})(q^{2}+\lambda^{2})}},
\end{eqnarray}
where $f(y)$ is a dimensionless function of the plasma density distribution against the height $y$ in the vertical direction of the accretion disc, which is related to the mass-density distribution $\rho(y)$ as $\rho(y)=n_{\rm e}m_{\rm H}f(y)$.

\begin{figure}
\centering
\includegraphics[width=0.45\textwidth]{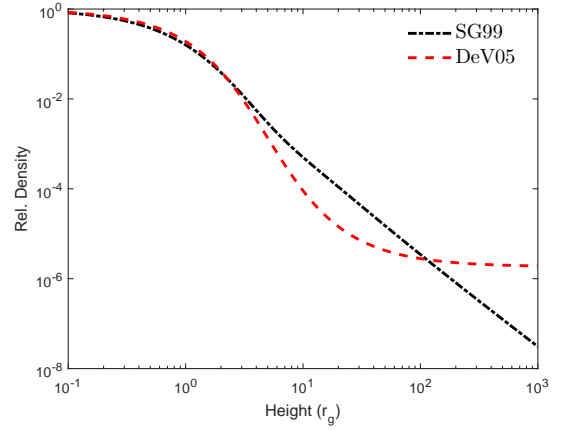}
\caption{Normalized density profile for two models. The black dot-dashed line (SG99) is for the solar atmosphere, while the red dashed line (DeV05) is from the three-dimensional numerical simulation of the hot accretion flow. See the text for details.}\label{fig:den}
\end{figure}

There are two choices for the density distribution $f(y)$. In the previous numerical models (e.g., \citealt{Yuan09,Meng15}), they adopted the solar model, in which $f(y)$ followed the empirical S\&G atmosphere (e.g., \citealt{Sittler99,Lin06}). In the present work, we adopt a more realistic density distribution in the corona which is taken from a fully general relativistic three-dimensional MHD simulation of hot accretion flows \citep{DeVilliers05}. They found that the number density decreased exponentially with decreasing polar angle (see Figure 3 of that paper).
As shown in Figure~\ref{fig:den}, the two density profiles differ slightly at the large height, but behave similarly in the lower latitude. Our tests show that the final results for the light curve modeling are insensitive to the two different density profiles. Therefore, we will present only the results from adopting the density profile from the numerical simulations \citep{DeVilliers05}.

Now we are ready to investigate the dynamical properties of the system following the catastrophe by solving differential
Equations~(\ref{eq:dh2dt}, \ref{eq:dpdt}, \ref{eq:dqdt}, \ref{eq:dmdt}) and $dh/dt=\dot{h}$.  The dynamical properties of the flux rope are described by five physical quantities ($p, q, h, \dot{h}, m$), which are the bottom, top tip of the current sheet, the height and velocity of flux rope, and the mass inside the flux rope per unit length, respectively. For the dynamical evolutions of the system, four free parameters are involved, namely the magnetic field strength $B_{0}$, the electron number density $n_{\rm e}$, the density contrast ratio for the flux rope $\xi$, and the Alfv\'{e}n Mach number $M_{\rm A}$ of the reconnection inflow.

\subsection{Energetics}\label{sec:engergy}
Based on the Poynting's theorem, the change rate of the thermal energy is equal to the  integral of the Poynting flux along the current sheet $S(t)$, which is a part of the magnetic energy \citep{Reeves05}. It is this part of magnetic energy that contributes to the observed radiation in the eruption. By calculating the Poynting flux in the current sheet, we obtain the power related to the energy dissipated in the current sheet. Specially, given $p,q,h$, and $\dot{h}$ as a function of time, we calculate the power associated with the dynamical evolution, which is \citep{Reeves06,Meng14}
\begin{eqnarray}  \label{eq:st}
 \frac{dW_{\rm EM}}{dt}&=&S(t) \nonumber \\
 &=&\frac{c}{2 \pi} E_{z}(t) \int^{q(t)}_{p(t)}B_{y}(0,y,t)dy,
\end{eqnarray}%
where $E_{z}(t)$ is the electric field in the reconnection region induced in the reconnection process \citep[see details given by][]{LF00}, the magnetic field along the current sheet $B_{y}(0,y,t)$ is determined by Equation~(\ref{eq:Bxy}) with  $x=0$, $q$ and $p$ are the top and bottom tips of the current sheet, respectively, as shown in Figure~\ref{fig:diagram}.  Here the product of the electric field $E_{z}$ and the magnetic field $B_{y}$ gives the Poynting flux that describes the electromagnetic energy flux entering the current sheet with the reconnection inflow.

Substituting Equations (\ref{eq:Bxy}) and (\ref{eq:Ez}) into Equation (\ref{eq:st}) and integrating, we have
\begin{eqnarray}\label{eq:Sst}
S\left( t \right)&=&\frac{V_{\rm A}(0,y_{0})}{{2\pi }}{\left( {\frac{{2{I_0}}}{c}} \right)^2}\frac{{{M_{\rm A}}{\lambda ^2}{{({h^2}
  + {\lambda ^2})}^2} }}{{q({p^2} + {\lambda ^2})({q^2} + {\lambda ^2})}}  \nonumber \\
 &\times&\frac{\sqrt {({y_0}^2 - {p^2})({q^2} - {y_0}^2)}}{({h^2} - {y_0}^2)({y_0}^2 + {\lambda ^2})}\left[K\left(\frac{{\sqrt {{q^2} - {p^2}} }}{q}\right) \right. \nonumber \\
   &- & \frac{{{p^2} + {\lambda ^2}}}{{{h^2} + {\lambda ^2}}}\Pi \left(\frac{{{q^2} - {p^2}}}{{{q^2} + {\lambda ^2}}},
  \frac{{\sqrt {{q^2} - {p^2}} }}{q}\right)   \nonumber \\
  &- & \frac{{{h^2} - {p^2}}}{{{h^2} + {\lambda ^2}}}\Pi \left(\frac{{{q^2} - {p^2}}}{{{q^2} - {h^2}}},
  \left.\frac{{\sqrt {{q^2} - {p^2}} }}{q}\right) \right],
\end{eqnarray}
where $K$ and $\Pi$ are the complete elliptic integral of the first kind and the third kind, respectively.

Note that $S(t)$ above is the power released in the reconnection process per unit length. By introducing the third dimensional length scale of the radiative process, i.e., the length of the flux rope, the energy injection rate powering the radiation in the flare regions is (refer to the left plot of Figure 1 in \citealt{Yuan09})
\begin{equation}\label{eq:Et}
\dot{E}(t)=\pi L_{0}S(t),
\end{equation}
where $L_{0}$ is the distance between two foot points of the flux rope anchored in the accretion flow. The particle inflow rate in the current sheet is related to the local electron number density $n_{\rm e}$ and the Alfv\'{e}n speed $V_{\rm A}$, which is expressed as
\begin{equation}\label{eq:Nt}
\dot{N}_{\rm th}(t)=2\pi M_{\rm A}V_{\rm A}[q(t)-p(t)]n_{\rm e}L_{0}
\end{equation}

We note that an extra physical parameter $L_{0}$ is added to calculate the power associated with the dynamical evolutions.

\subsection{Injected Electron Distribution}\label{sec:dis}

With the energy and particle injection rates in hand, we now discuss the initial energy spectrum of electrons injected in the flare region at a function of time. We consider a hybrid distribution of electrons, i.e., the electrons are in the mixture of both thermal and power-law distributions \citep{Yuan03}. The (relativistic) thermal distribution with the total normalized particle number per unit time $\dot{N}_{\rm th}(t)$ is
\begin{equation}
 n_{\rm th}(\gamma)={\dot{N}_{\rm th}(t)\gamma^2\beta \exp (-\gamma/\theta_{\rm e}) \over
\theta_{\rm e} {K}_2(1/\theta_{\rm e})},
\end{equation}
where $\gamma=1/\sqrt{1-\beta^2}$ is the electron Lorentz factor, $\theta_{\rm e}\equiv kT_{\rm e}/m_{\rm e} c^2$ is the dimensionless
electron temperature, and $K_2$ is the modified Bessel function of the second order.

The power-law distribution is described by
\begin{equation}
n_{\rm pl}(\gamma)=c_{\rm inj}'\gamma^{-p_{\rm e}}, \hspace{1cm}
 \gamma_{\rm min} \le
\gamma \le \gamma_{\rm max},
\end{equation}
where $\gamma_{\rm min}$ and $\gamma_{\rm max}$ are the minimum and maximum Lorentz factors of electrons, respectively, $c_{\rm inj}'$ is the power-law normalization.

We calculate the values of $\gamma_{\rm min}$ and $c_{\rm inj}'$ as follows.  We assume that the {\it injected} energy in non-thermal electrons is equal to a fraction $\eta$ of the energy in thermal electrons. We further assume that $\eta$ is small and independent of time.  The injected energy flux at a given time $t$ of thermal electrons at temperature $\theta_{\rm e}$ is \citep{Chandrasekhar39}

\begin{equation}\label{eq:uth}
 u_{\rm th}(t)=a(\theta_{\rm e})\dot{N}_{\rm th}(t)m_{\rm e}c^2\theta_{\rm e},
\end{equation}
where the quantity
\begin{equation}
a(\theta_{\rm e})\equiv\frac{1}{\theta_{\rm e}}\left[\frac{3K_3(1/\theta_{\rm e})+K_1(1/\theta_{\rm e})}
{4K_2(1/\theta_{\rm e})}-1\right]
\end{equation}
varies from $3/2$ for nonrelativistic
electrons to $3$ for fully relativistic electrons, and $K_{\rm n}$ are modified Bessel functions of the $n$th order. The electron temperature is determined by setting $u_{\rm th} (t) = \dot{E}(t)$ in Equation~(\ref{eq:Et}), which can now be expressed as
\begin{equation}
\dot{E}(t)=a(\theta_{\rm e})\dot{N}_{\rm th}(t)m_{\rm e}c^2\theta_{\rm e}.
\end{equation}

The energy density of power-law electrons is
\begin{equation} \label{eq:upl}
u_{\rm pl}\approx \frac{c_{\rm inj}'}{p_{\rm e}-2}m_{\rm e}c^2\gamma_{\rm min}^{2-p_{\rm e}}
\end{equation}
for $p_{\rm e} > 2$. So the normalization of power-law electrons is determined by $u_{\rm
pl}=\eta u_{\rm th}$, which gives
\begin{equation} \label{eq:Npl}
c_{\rm inj}'(t)=(p_{\rm e}-2)\gamma_{\rm min}^{p_{\rm e}-2}\eta a(\theta_{\rm e})
\theta_{\rm e}\dot{N}_{\rm th}(t).
\end{equation}
 If $p_{\rm e} < 2$, the formula corresponding to Equations~(\ref{eq:upl}) and (\ref{eq:Npl}) are
\begin{equation} \label{eq:upl2}
u_{\rm pl}\approx \frac{c_{\rm inj}'}{2-p_{\rm e}}m_{\rm e}c^2\gamma_{\rm max}^{2-p},
\end{equation}
and
\begin{equation}\label{eq:Npl2}
c_{\rm inj}'=(2-p_{\rm e})\gamma_{\rm max}^{p_{\rm e}-2}\eta a(\theta_{\rm e}) \theta_e\dot{N}_{\rm th}.
\end{equation}

Another constraint is that the power-law distribution should smoothly match the thermal distribution at $\gamma_{\rm min}$,
\begin{equation} \label{eq:ngmin}
n_{\rm th}(\gamma_{\rm min})=n_{\rm pl}(\gamma_{\rm min}).
\end{equation}
This condition is naturally expected since the non-thermal electrons are
presumably accelerated out of the thermal pool.
We can then calculate numerically $c_{\rm inj}'$ and $\gamma_{\rm min}$  as a function of time by solving Equations~(\ref{eq:Npl}) and (\ref{eq:ngmin}) simultaneously.

The value of $\gamma_{\rm max}$ depends on the details of electron acceleration which are not well understood. We treat it as a constant $\gamma_{\rm max}=10^{6}$. Note that the exact value of $\gamma_{\rm max}$ (as long as being large enough) is unimportant for the X-ray and NIR flares considered here.

The outflowing particles from the current sheet will flow into two different flare regions, the flux rope and the flare loop as we show in Fig. 1. A reasonable assumption is that half of $c_{\rm inj}'$ goes into each of them, namely $c_{\rm inj,rope}=c_{\rm inj,loop}=1/2c_{\rm inj}'$. This assumption is equivalent to $\dot{E}_{\rm rope}(t)=\dot{E}_{\rm loop}(t)=1/2\dot{E}(t)$ and $\dot{N}_{\rm th,rope}(t)=\dot{N}_{\rm th,loop}(t)=1/2\dot{N}_{\rm th}(t)$. In the following numerical modeling, we define $c_{\rm inj}(t)$ as  the injection profile for both the flux rope and the flare loop region to avoid confusions.

What are the values of  $p_{\rm e}$ and $\eta$? Observations of solar flares have revealed a high particle energization efficiency, i.e., $10\%-50\%$ of the magnetic energy ejected into power-law particles  \citep{Lin76}. As a fiducial model, we adopt $\eta=0.1$ as a fixed value.  The value of $\eta$ is not very important as it can be partly absorbed by $L_{0}$. The remaining important parameter is $p_{\rm e}$. We will show below that many theoretical and simulation works will also constrain this power-law index of the electrons accelerated by magnetic reconnection.

\subsection{Light Curve and SED}\label{sec:lc}

We follow the method in \citet{DoddsEden10} to calculate the model light curves and spectra. The electron distribution function $N_{\rm e}(\gamma,t)$ (the number of electrons with  Lorentz factor $\gamma$ at time $t$) evolves according to the following continuity equation (\citealt{Blumenthal70})

\begin{equation}\label{eq:continuity}
  \frac{\partial{N_{\rm e}(\gamma,t)}}{\partial{t}}= Q_{\rm inj}(\gamma,t)-\frac{\partial[\dot{\gamma}N_{\rm e}(\gamma,t)]}{\partial{t}}-\frac{N_{\rm e}(\gamma,t)}{t_{\rm esc}(\gamma,t)}.
\end{equation}
The escape term $t_{\rm esc}(\gamma, t)$ can be described by the diffusive escape timescale from the system, or by the timescale for catastrophic losses, such as those which occur in the extreme Klein-Nishina limit for electrons, or in secondary nuclear processes for hadronic collisions (\citealt{Dermer09}). Without addressing these microphysics in detail, we simply assume $t_{\rm esc}=1000~\rm mins$ in this work, which is much longer than the typical flare timescales for \sgra.

When $t_{\rm esc}(\gamma,t)\rightarrow t_{\rm esc}(\gamma)$ and $\dot{\gamma}<0$, Equation~(\ref{eq:continuity}) has a solution as (\citealt{Blumenthal70,Dermer09}, see their Appendix C)
\begin{eqnarray}\label{eq:nev}
  \nonumber N_{\rm e}(\gamma,t)&=& \frac{1}{|\dot{\gamma}|}\int_{\gamma}^{\infty}d\gamma^{\prime}{Q_{\rm inj}}(\gamma^{\prime},t^{\prime}) \\
  &&  \times\exp\left(-\int_{\gamma}^{\gamma^{\prime}}\frac{1}{t_{\rm esc} (\gamma^{\prime\prime})}\frac{d\gamma^{\prime\prime}}{|\dot{\gamma}|} \right),
\end{eqnarray}
where
\begin{equation}
t^\prime=t-\int_{\gamma}^{\gamma\prime}\frac{d\gamma^{\prime\prime}}{|\dot{\gamma}|},
\end{equation}
and $Q_{\rm inj}(\gamma,t)$ is the rate at which electrons with the Lorentz factor $\gamma$ are injected at time $t$ and can be taken as a power law in $\gamma$: $Q_{\rm inj}(\gamma,t)=c_{\rm inj}(t)\gamma^{-p_{\rm e}}$.

We consider two cooling processes for $\dot{\gamma}$ in Equation~(\ref{eq:continuity}), synchrotron and adiabatic cooling.  For the case of synchrotron cooling, we have
\begin{equation}\label{eq:gdot:syn}
\dot{\gamma}_{\rm syn}=-\gamma/t_{\rm syn},
\end{equation}
where $t_{\rm syn}=7.7462\times10^8/(\gamma{B^2})~{\rm s}$ for an isotropic pitch angle distribution, and $B$ is in units of Gauss. For the two flare regions, the magnetic field profiles are different, both of which can be determined by Equation~(\ref{eq:Bxy}). We adopt a spatial averaged magnetic field profile in order to simplify the numerical model. For the erupted flux rope region (refer to Fig. 1), it is expressed as
\begin{equation}\label{eq:brope}
B_{\rm rope}(t)=\left\langle B(x=0,q\leq y\leq h)\right\rangle
\end{equation}
where $\langle\rangle$ means the average over $x=0, q\leq y\leq h$. While for the flare loop below the current sheet (refer to Fig. 1), the magnetic filed is averaged over the region $x=0$ and $y\leq p$,
\begin{equation}\label{eq:bloop}
B_{\rm loop}(t)=\left\langle B(x=0,0\leq y\leq p)\right\rangle .
\end{equation}
The adiabatic cooling rate is
\begin{equation}\label{eq:gdot:ad}
\dot{\gamma}_{\rm ad}=-\gamma~{d\log{R}}/{dt}=-\gamma~v_{\rm exp}/R,
\end{equation}
where the expansion velocity $v_{\rm exp}=d{R}/dt$.  The adiabatic cooling rate is also different for different flare regions, depending chiefly on the expansion behavior of the radiative blob. For the ejected blob enclosing the flux rope region as shown in Figure~\ref{fig:diagram}, $R\simeq{h-q}$, while for the flare loop close to the surface of accretion flow, $R\simeq{p}$.

The total cooling rate is thus
\begin{equation}\label{eq:gdot}
\dot{\gamma}=\dot{\gamma}_{\rm syn}+\dot{\gamma}_{\rm ad}.
\end{equation}
The synchrotron emission is calculated at each time given the instantaneous electron energy distribution using the following formulae \citep{Rybicki79}:
\begin{equation}\label{eq:jv}
  j_{\nu}=\frac{1}{4\pi}\int_1^{\infty}n_{\rm e}(\gamma)\langle{P_{\rm e}(\gamma,\nu,\theta)}\rangle d\gamma,
\end{equation}
where $\langle{P_{\rm e}(\gamma,\nu,\theta)}\rangle$ is the pitch angle (LOS and B) averaged spectral power emitted by a single electron and
\begin{equation}
P_{\rm e}(\gamma,\nu,\theta)=\frac{\sqrt{3}e^3B\sin\theta}{mc^2}F\left(\frac{\nu}{\nu_{\rm syn}(\gamma,\theta)}\right),
\end{equation}
with
\begin{equation}
\nu_{\rm syn}(\gamma,\theta)=3eB\gamma^2\sin\theta/(4\pi{mc})
\end{equation}
and
\begin{equation}
F(x)=x\int_x^{\infty}K_{5/3}(\xi)d\xi.
\end{equation}
The absorption coefficient is
\begin{equation}\label{eq:lvabs}
  \alpha_{\nu} = \frac{c^2}{8\pi\nu^2mc^2}\int_1^{\infty}n_{\rm e}(\gamma)\left(\frac{2P_{\rm e}(\gamma)}{\gamma}+\frac{dP_{\rm e}(\gamma)}{d\gamma}\right)d\gamma.
\end{equation}
Assuming a homogeneous sphere of radius $R$ the resultant emission is \citep{Gould79,DoddsEden10}
 \begin{eqnarray}\label{eq:lvsyn}
   \nonumber \nu L_{\nu} &=& 4\pi^2R^2\frac{\nu j_{\nu}}{\alpha_{\nu}}\\
   &\times&\left(1+\frac{\exp(-2\alpha_{\nu} R)}{\alpha_{\nu}{R}}-\frac{1-\exp(-2\alpha_{\nu}R)}{2\alpha_{\nu}^2R^2}\right).
 \end{eqnarray}
Note that if the emission is optically thin, the luminosity only depends on the total number of accelerated electrons $N_{\rm e}(\gamma,t)= 4\pi /3 R^3 n_{\rm e}(\gamma,t)$, so Equation~(\ref{eq:lvsyn}) can be reduced to
\begin{eqnarray}\label{eq:Lvthin}
  \nonumber \nu L_{\nu} &=& 4\pi \frac{4\pi R^3}{3}\nu{j_{\nu}}\\
  &=&\nu\int_1^{\infty}N_{\rm e}(\gamma,t)\langle{P_{\rm e}(\gamma,\nu,\phi)}\rangle d\gamma.
\end{eqnarray}


\section{Numerical Results}\label{sec:result}

\subsection{Fiducial Model}


Our model has six parameters, namely $B_{0}$, $n_{\rm e}$, $\xi$, $M_{\rm A}$, $L_{0}$, and $p_{\rm e}$. The model can produce light curves and SED of the flares. The characteristic values of several important parameters can be constrained from either the observations of \sgra\ or theoretical works, thus only leave limited room for adjustment.

All spatial lengths of the system are scaled by $r_{\rm g}\equiv GM_{\bullet}/c^2$, the gravitational radius of the black hole. The half-distance of two foot points of the magnetic loop is chosen to be $\lambda_{0}=5r_{\rm g}$ according to \citet{Meng15}. The flux rope radius is hard to estimate and adopted to be $r_{00}=0.1\lambda_{0}=0.5r_{\rm g}$ following the CME model for our Sun \citep{LF00}. For the strength of magnetic field in the accretion flow of \sgra, many authors suggest that $B\lesssim30~\rm G$ (e.g., \citealt{Yuan03,Sharma07,DoddsEden10} and references in the Introduction section). In a localized flare region, however, the field could be much stronger.

We can also reasonably estimate the value of the number density of electrons in the accretion flow, $n_{\rm e}$, from previous observational and theoretical works.  \citet{Yuan03} found that the number density in the equatorial plane of the innermost region of the accretion flow is $n_{\rm e}\sim10^{7}~{\rm cm^{-3}}$ by modeling the quiescent spectrum of \sgra. Observationally, we have good constraints on the number density at the Bondi radius by \emph{Chandra} observation \citep{Baganoff03,Wang13}. Numerical simulations of the hot accretion flow covering four orders of magnitude in the radial dynamical range by \citeauthor{Yuan12b}(\citeyear{Yuan12b}; see also references therein) has shown that the radial density distribution of the accretion flow can be well described by a power-law form,  $n_{\rm e}\propto r^{-s}$, with the index $s$ in the range of $0.5-1.0$. Combining this result with the observations of \emph{Chandra}, we can obtain the number density in the inner region of the accretion flow for \sgra, which is also close to $n_{\rm e}\sim10^{7}~{\rm cm^{-3}}$.

The rates of energy release during the rise and decay phases in the process of magnetic reconnection are controlled by two different physical processes. The rise phase is mainly driven by an ideal-MHD process, which is determined by the Alfv\'{e}n timescale. After the loss of the equilibrium, a current sheet will be formed to halt further evolution of the system, unless reconnection starts. In the decay phase where reconnection begins to dominate in the dynamical evolution, the evolution is determined by the reconnection timescale. Thus, the Alfv\'{e}n Mach number $M_{\rm A}$ can determine the ratio of rise and decline timescales, namely the asymmetry of flare profiles.

In the Sweet-Parker model of the magnetic reconnection, the reconnection speed is only a tiny fraction of Alfv\'{e}n speed, which means the Alfv\'{e}n Mach number $M_{\rm A}$ is extremely smaller than unity. However, observations in solar flares require the reconnection speed close to the Alfv\'{e}n speed. One way to speed up the reconnection is to invoke plasma instabilities, for example, the stream instability which makes Ohmic magnetic resistivity anomalously large \citep{Parker79}. Another way is to consider the presence of turbulence in the current sheet \citep{Lazarian99,delValle16}. We adopt $M_{\rm A}=0.5$ as to make the timescale of the rise phase comparable to the decay phase of the flares, which corresponds to the case of quasi-symmetric flares.

The energy spectral index of accelerated electrons is complicated to determine, and depends on the initial and the boundary conditions of the reconnection site (see review by \citealt{deGouveia15}). Analytical studies of the first order Fermi process in current sheets predict that the power-law index $p_{\rm e}=2.5$ \citep{deGouveia05} or $1.0$ \citep{Drury12}. However, \citet{Kowal12} found a hard power-law spectrum with $p_{\rm e}=1.0$ for particle acceleration in 3D MHD reconnection sites, close to $p_{\rm e}\sim1.5$ obtained from 2D collisionless PIC simulations considering merging islands \citep{Drake10}. These uncertainties provide us with some flexibilities to choose the value of $p_{\rm e}$.

\begin{figure}
\centering
\includegraphics[width=0.45\textwidth]{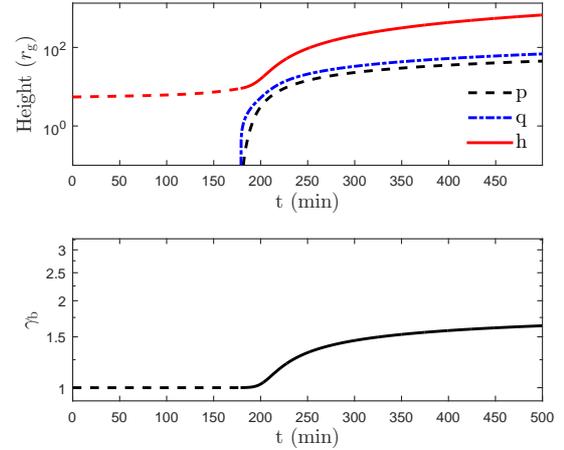}
\caption{Dynamics of the flux rope.
Upper panel: the red solid, blue, and black lines represent the evolutions of the height of the flux rope $h$, and the current sheet ($q, p$) as a function of time.  The red dashed line corresponds to the evolution prior to the loss of the equilibrium.  Bottom panel: the evolution of the Lorentz factor $\gamma_{\rm b}$ of the flux rope. Since the magnetic reconnection can only take place in the last stage, which is after the loss of equilibrium, our calculations in the following are mainly focused in the last stage. }\label{fig:dyn}
\end{figure}

As for the length of the flux rope, the observational constraint is less certain, and we adopt $L_{0}=50r_{\rm g}$,  the same order of magnitude as the height of the flux rope (see Figure~\ref{fig:dyn}). As we will see, the impact of $L_{0}$ is simply only to affect the magnitude of the flare luminosity in NIR and X-ray in a same way.

Accordingly, we choose the characteristic values of these parameters mentioned above as summarized in Table~\ref{tab:parameters}.

Solving differential Equations~(\ref{eq:dh2dt}, \ref{eq:dpdt}, \ref{eq:dqdt}, \ref{eq:dmdt}) and $dh/dt=\dot{h}$ as shown in Section~\ref{sec:dyn} will give the dynamical evolution of the system following the catastrophe. The initial conditions for catastrophe are given by $h(t = 0)$, $\dot{h}(t = 0)$, $p(t =0)$, $q(t = 0)$ and $m(t = 0)=m_{0}$, which control the evolution in the first stage. The determination of $h(t = 0)$, $\dot{h}(t = 0)$, $p(t =0)$, $q(t = 0)$ can be found in \citet{LF00}, while $m_{0}$ is determined by our model parameters as discussed in Section~\ref{sec:dyn}. We show the results as the dashed lines in Figure~\ref{fig:dyn}. As it is an ideal MHD process without magnetic energy release to power the radiative flare in this stage, we don't duplicate the calculations here and refer the interested readers to \citet{LF00} for details.

\begin{figure}
\centering
\includegraphics[width=0.45\textwidth]{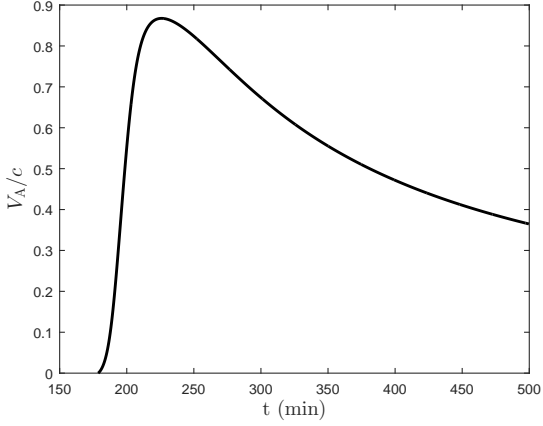}
\caption{Evolution of local Alfv\'{e}n speed at the centre of the current sheet $(0,y_{0})$ in units of the speed of light. }\label{fig:VA}
\end{figure}

The dynamical evolution of the system is shown as the solid lines in Figure~\ref{fig:dyn}. It is clear that the flux rope can be accelerated to mid-relativistic speed within several minutes and the height of the flux rope can be as high as several hundred $r_{\rm g}$. We can see that the top and bottom ends of the current sheet are very close to each other, as indicated by the blue dot-dashed and black dashed lines in the upper panel of the figure. This closeness mainly arises from a large Alfv\'{e}n Mach number $M_{\rm A}=0.5$ that we have adopted, which makes the current sheet reconnect very efficiently. These profiles are quantitatively similar to (although not as dramatic as) the results presented in \citet{Yuan09}. The velocity profile of the flux rope as shown in the bottom panel of Figure~\ref{fig:dyn} is much shallower compared to the dramatic eruption profile in \citet{Yuan09} due to the fact that the Alfv\'{e}n timescale inferred there is longer than that in this work. In Figure~\ref{fig:VA}, we present the evolution of the Alfv\'{e}n velocity.

\begin{figure}
\centering
\includegraphics[width=0.45\textwidth]{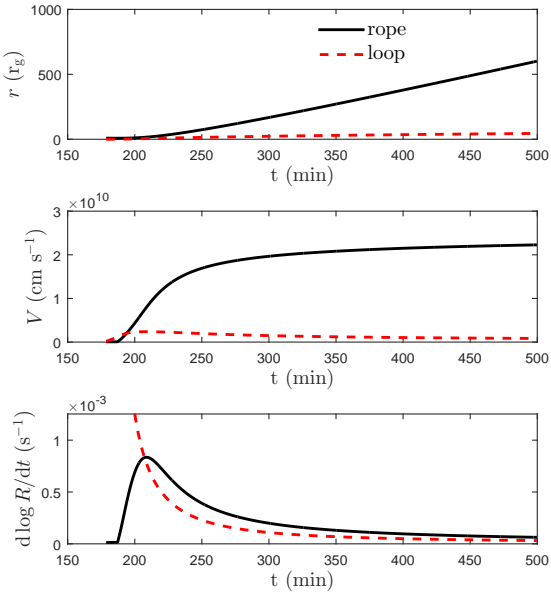}
\caption{Adiabatic expansion of the flux rope (solid lines) and flare loop (dashed lines) regions. Upper panel: the size of the two flare regions. Middle panel: the expansion velocity of the two flare regions considered in the upper panel. Lower panel: the adiabatic cooling term  in Equation~(\ref{eq:gdot:ad}) for the two flare regions considered above.}\label{fig:ad}
\end{figure}

To obtain the adiabatic cooling term in modeling the time-dependent distribution of injected electrons, we calculate the term $d\log{R}/dt$ (refer to Equation~(\ref{eq:gdot:ad}), as shown in Figure~\ref{fig:ad}. There are two flare regions as discussed in Section~\ref{sec:dis}, i.e., the flux rope and flare loop shown in Figure~\ref{fig:diagram}. We calculate the corresponding term in these two regions with the method presented in Section~\ref{sec:dis}. As shown in Figure~\ref{fig:ad}, the  expansion velocity of the size of ``flux rope'' region is much larger than that of the ``flare loop'' region. This implies a fast cooling rate in the late stage of the evolution, as shown in the bottom panel of Figure~\ref{fig:ad}. Note that in the beginning of the catastrophe, the cooling rate in the loop region is  larger than that of the flux rope. This is partly owing to the much smaller size of the flare loop region in the initial stage following the loss of equilibrium when the bottom end of the current sheet is still close to the accretion flow surface.

\begin{figure}
\centering
\includegraphics[width=0.45\textwidth]{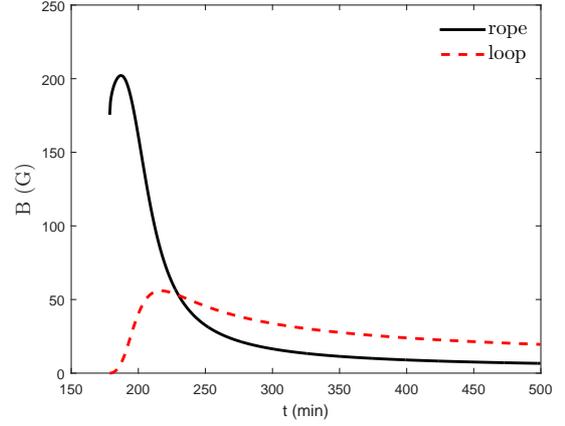}
\caption{Evolution of the magnetic field in the flux rope (solid line) and flare loop (dashed line) regions. A spatially averaged value is adopted according to Equation~(\ref{eq:brope}) and (\ref{eq:bloop}) to simplify the time-dependent modeling.}\label{fig:magB}
\end{figure}

Another important factor determining the cooling rate and the consequent radiation is the distribution of magnetic field in the flare regions. The evolution of the spatially averaged magnetic field in the ``flare loop'' and ``flux rope'' regions are shown in Figure~\ref{fig:magB} based on Equations~(\ref{eq:bloop}) and (\ref{eq:brope}). It is clearly shown that the magnetic field strength decreases significantly after magnetic reconnection starts. It is surprising that there is a rapid increase initially in the magnetic field in the loop region. We can also see that the maximum field strengths are different in the two different regions. The field strength in the flux rope region can be higher than $B_{0}$ set in Table~\ref{tab:parameters}. This is because the magnetic flux accumulates as the magnetized plasma flowing into the flare regions.

\begin{figure}
\centering
\includegraphics[width=0.45\textwidth]{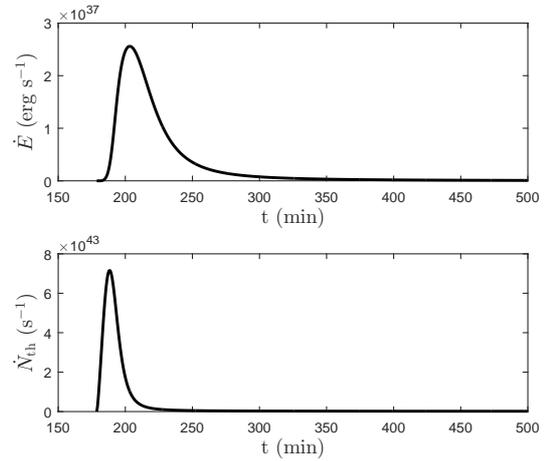}
\caption{Evolution of the total (including flux rope and flare loop regions) energy release rate (upper) and particle injection rate (bottom). }\label{fig:Edot}
\end{figure}

We further discuss the energy release during the eruption process. The power output associated with the dynamical evolution is directly related to $\dot{E}(t)$ in Equation~(\ref{eq:Et}). The shape of $\dot{E}(t)$ shown in Figure~\ref{fig:Edot} is reminiscent of typical light curves of flares in NIR and X-ray. The quasi-symmetric profile of $\dot{E}(t)$ is due to the relative large value of $M_{\rm A}=0.5$, which results in a steep (or soft) tail in the decline phase of the light curve. Assuming that all the energy dissipated can be converted to radiative flares, \citet{Meng14} used $\dot{E}(t)$ to compare with the observed light curve. This simplification could overestimate the efficiency of the radiation. As shown in the upper panel of Figure~\ref{fig:Edot}, the energy release rate is about 2 orders of magnitude higher than the peak luminosity of typical flares as shown in Figure~\ref{fig:lc}, indicating a radiative efficiency of only $\sim1\%$. In the bottom panel of Figure~\ref{fig:Edot}, we calculate the electron number flowing into the current sheet. A fraction of which is accelerated into a power-law distribution by the magnetic reconnection. The total particle injection profile $\dot{N}(t)$ is slightly narrower than $\dot{E}(t)$, but is also relatively symmetric.

\begin{figure}
\centering
\includegraphics[width=0.45\textwidth]{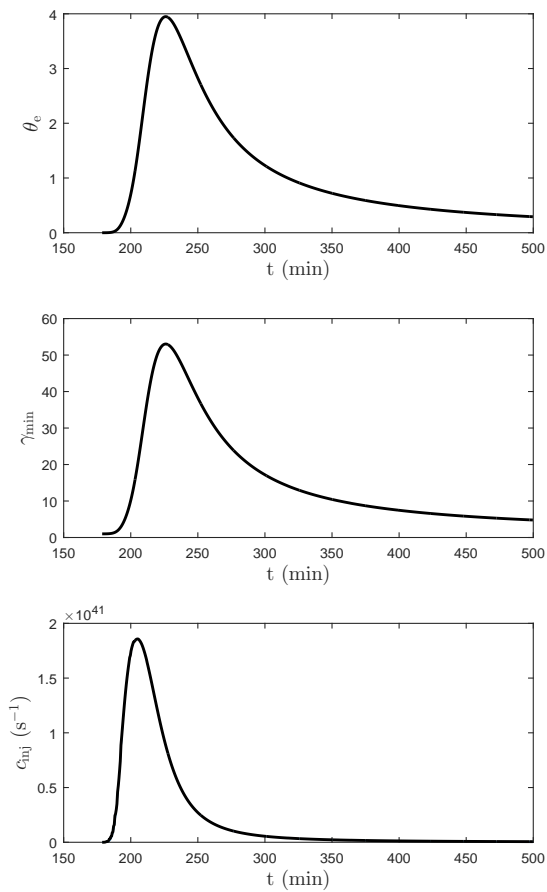}
\caption{Time evolution of the total (including flux rope and flare loop regions) injected electron properties. Upper panel: the dimensionless temperature. Middle panel: the minimum Lorentz factor of power-law electrons. Bottom panel: the normalization of the power-law distribution. }\label{fig:cinj}
\end{figure}

With the time profile of energy injection rate and the particle injection rate, we can now obtain the time evolution of various physical quantities, namely the dimensionless electron temperature $\theta_{\rm e}(t)$, the minimum Lorentz factor of power-law electrons $\gamma_{\rm min}(t)$ and the injection profile $c_{\rm inj}(t)$, as described in Section~\ref{sec:dis}. We show the numerical results in Figure~\ref{fig:cinj}. Note that $\dot{N}(t)$ is the total electron number by integrating over $\gamma$, while $c_{\rm inj}$ represents the normalization of the power-law electrons.  It is the injection profile $c_{\rm inj}$ that determines the shape of the resulting light curves. As expected, we can see that the shape, including the rise, decay phase and the width of the light curve,  resembles the typically observed X-ray light curves.

\begin{figure}
\centering
\includegraphics[width=0.45\textwidth]{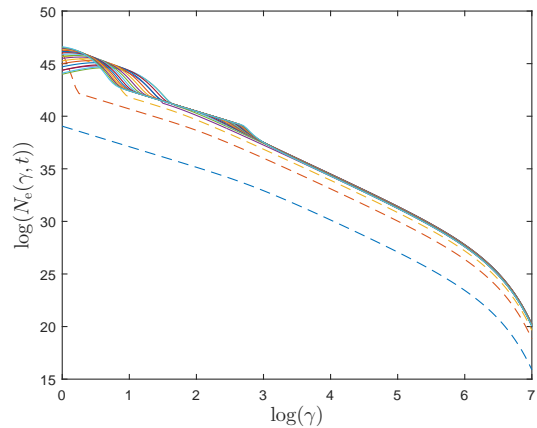}
\caption{Time evolution of the electron energy spectrum. The dashed lines correspond to the rise phase, the solid lines are for the decay phase. The spectra are calculated for every 20~min.}\label{fig:Ne}
\end{figure}

\begin{figure}
\begin{center}
\includegraphics[width=0.45\textwidth]{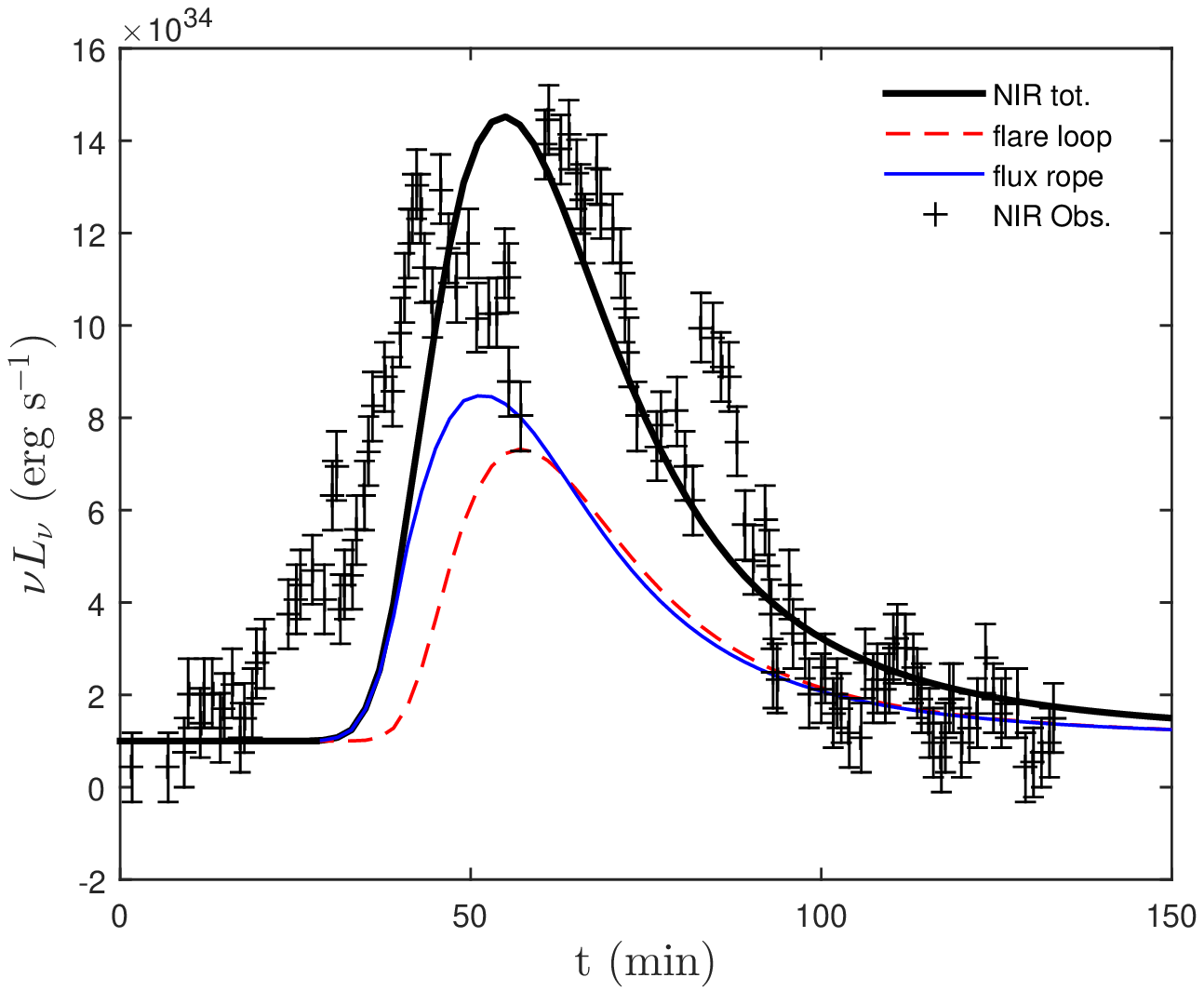}
\includegraphics[width=0.45\textwidth]{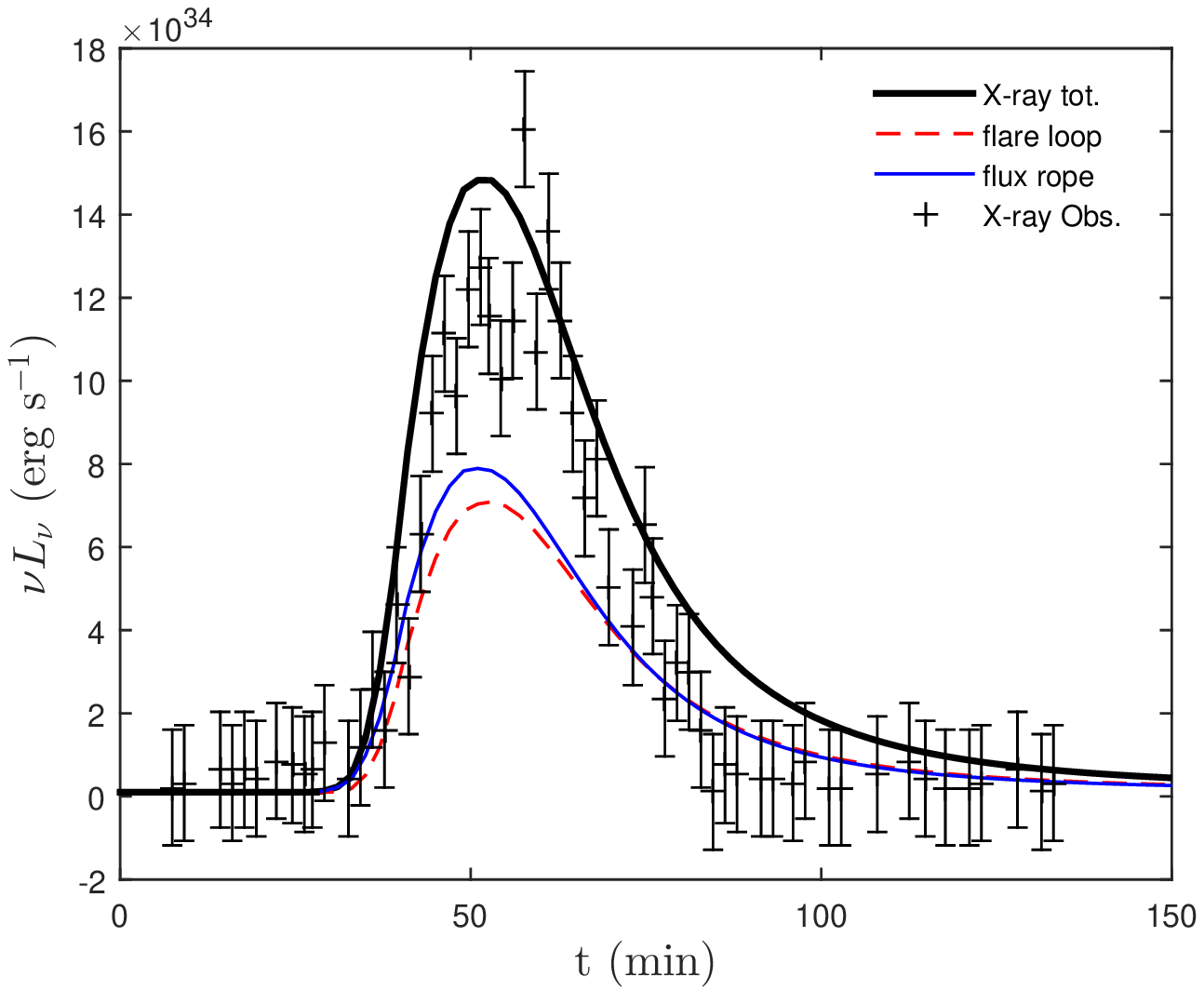}
\end{center}
\caption{NIR and X-ray flare light curves of \sgra. Upper panel: the blue solid and red dashed line correspond to the emission from flux rope and flare loop regions, respectively. The theoretical NIR light curve represented by the thick black line is the sum of the two components. For comparison, the black points with error bars represent the observational data taken on April 4, 2007 \citep{DoddsEden09}. Bottom panel: theoretical X-ray light curve and the observed one during the same period. }\label{fig:lc}
\end{figure}

With the injection profile $c_{\rm inj}(t)$ and the minimum Lorentz factor $\gamma_{\rm min}(t)$ in hand, we can obtain the injection term $Q_{\rm inj}(\gamma,t)$ in Equation~(\ref{eq:continuity}). The cooling term $\dot{\gamma}$ can also be modeled with two terms discussed above. The adiabatic term is shown in Figure~\ref{fig:ad}, while the synchrotron term can be determined by the magnetic field profile in the corresponding flare regions. We can then solve the continuity equation in Section~\ref{sec:lc} to obtain the time evolution of the energy spectrum of electrons. The results are shown in Figure~\ref{fig:Ne}. A broken power-law feature in the electron spectra at $\gamma\sim 10^3$ exists in the rise phase. This is due to the short cooling timescale of electron compared to the injection timescale, which results in a spectral index of $p_{\rm e}'=p_{\rm e}+1$  \citep{Rybicki79}.

The NIR and X-ray emissions considered here are optically thin, which allows us to utilize Equation~(\ref{eq:Lvthin}) to directly calculate the emergent light curves and spectra. The results are shown in Figure~\ref{fig:lc}. In the upper panel of the figure, we compare the model light curve (solid line) to the NIR data (black points) taken on April 4, 2007. Since coordinated X-ray flare were observed, we also calculate the model light curve in X-ray based on synchrotron radiation using the same model parameters.
We can see that the X-ray light curve almost peaks simultaneously with the NIR, consistent with the observed data. Due to the fact that the cooling timescale of X-ray emitting electrons is much shorter than the injection timescale, the electrons radiate all of the energy supplied via injection. In that case, the shape of the X-ray light curve follows the injection profile very well, independent of $\gamma_{\rm min}(t)$ and $B(t)$.

As shown in Figure~\ref{fig:lc}, the emissions from the two flare regions, i.e.,flare loop and flux rope, are comparable for the NIR and X-ray flares. The total contribution at both two bands are thus the sum from these two flare regions. Since the flare loop makes a comparable contribution to the total emission in both NIR and X-ray, observationally these flares appear to be associated with expanding hot spots close to the black hole. Moreover, there are astrometric signatures during strong flares since we find that such flares are associated with the ejection of the magnetized plasmoids from the inner region of the accretion flow. Such expanding and/or ejected blobs could be detected by future high resolution instrument, such as Very Large Telescope Interferometer (VLTI) GRAVITY \citep{Eisenhauer11}. In addition, our preliminary analysis suggests that the emission from the flux rope region dominates over that from the flare loop for radio flares (Li et al. in preparation). Such a rapid expansion velocity for the radio-emitting (flux rope) region is actually consistent with current radio observations \citep[e.g.,][]{Brinkerink15}, and could be falsified by the near future Event Horizon Telescope (EHT) observations, which make the detection of the source size evolution during the sub-millimeter flares possible.

With the magnetic field strength and electron density used in this work, we find that the contribution from synchrotron self-Compton to the X-ray flares can be negligible, as argued in \citet{DoddsEden09}. It is also impossible to interpret the X-ray flares under reasonable physical parameters by inverse Compton process with seed photos from sub-millimeter emission. We thus neglect the Compton process for X-ray flares emission in this work.

\begin{figure}
\includegraphics[width=0.45\textwidth]{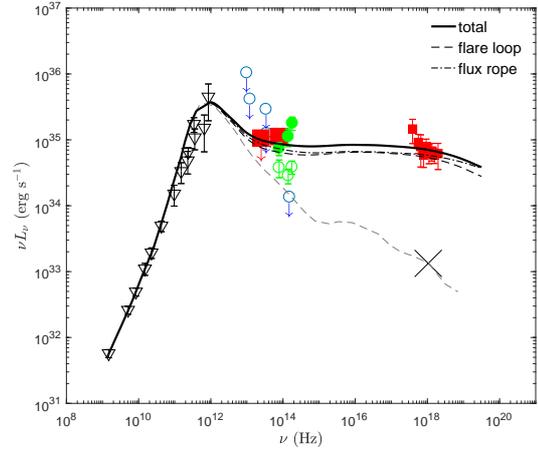}
\caption{Spectral modeling of the flares of \sgra. The observed data are quoted from \citet{DoddsEden09}. The filled red and green symbols present the flare state emission in infrared and X-ray (2-10 keV), other data points correspond to the quiescent emission.  The gray dashed line shows the RIAF model for the quiescent state emission taken from \citet{Yuan03}. The dashed and dot-dashed line correspond to the time-averaged (during flare) emission from flare loop and flare loop regions, respectively. The thick solid line is the total SED from our model.}
\label{fig:sed}
\end{figure}

With the time-dependent electron energy spectra, we can also calculate the SED of the flares. Here a time-averaged spectrum over the flaring period is calculated.
The result is shown in Figure~\ref{fig:sed}. For reference the quiescent model of \citet{Yuan03} is plotted (dashed gray line) and overlaid with the flare spectrum. The observed SED data are extracted from \citet{Markoff01,Baganoff03,Zhao03,Genzel03} for quiescent and \citet{Genzel03,DoddsEden09} for flare emissions. The thick solid line is the flare spectrum of \sgra. We can see that it is roughly consistent with observational data. However, we find that the X-ray spectra in our model are slightly harder than the observed one. This is mainly attributed to a high synchrotron cooling break frequency. Although the magnetic field strength is very high initially ($\sim100$ Gauss), it becomes rather weak during the flare peak ($\sim10$ Gauss) due to the reconnection (Figure~\ref{fig:magB}). As a result, the cooling break frequency is very close to the X-ray band. One solution is to increase the magnetic field strength, which would cause a more efficient cooling.

\begin{table*}
  \begin{center}
  \caption{Parameters for the MHD model of \sgra\ flares} \label{tab:parameters}
  \resizebox{0.75\textwidth}{!}{
  \begin{tabular}{ccccccccccc}
    \hline\hline
    model & $\lambda_0$ & $r_{00}$ & $B_0$  & $n_{\rm e}$ & $\xi$ & $M_{\rm A}$  & $L_{0}$ &  $p_{\rm e}$ & $\eta$ & $t_{\rm esc}$ \\
     & (1) & (2) & (3)  & (4) & (5) &(6) & (7) &  (8) & (9) & (10)\\

    \hline
     \texttt{fiducial} & $5$  & $0.5$   & 135 & $1.6\times10^7$  & $10$ & $0.5$    & 50        & 1.95            & 0.1  & $1000$    \\
     \hline
     \texttt{B} & $5$  & $0.5$   & 125 & $1.5\times10^7$  & $10$ & $0.5$    & 50        & 2.05            & 0.1  & $1000$    \\
       & $5$  & $0.5$   & 130 & $1.5\times10^7$  & $10$ & $0.5$    & 50        & 2.05            & 0.1  & $1000$   \\
       & $5$  & $0.5$   & 135 & $1.5\times10^7$  & $10$ & $0.5$    & 50        & 2.05            & 0.1  & $1000$   \\
       & $5$  & $0.5$   & 145 & $1.5\times10^7$  & $10$ & $0.5$    & 50        & 2.05            & 0.1  & $1000$   \\
     \hline
       \texttt{ne} & $5$  & $0.5$   & 135 & $1\times10^7$  & $10$ & $0.5$    & 50        & 2.05            & 0.1  & $1000$   \\
       & $5$  & $0.5$   & 135 & $1.5\times10^7$  & $10$ & $0.5$    & 50        & 2.05            & 0.1  & $1000$    \\
       & $5$  & $0.5$   & 135 & $1.8\times10^7$  & $10$ & $0.5$    & 50        & 2.05            & 0.1  & $1000$    \\
       & $5$  & $0.5$   & 135 & $2.2\times10^7$  & $10$ & $0.5$    & 50        & 2.05            & 0.1  & $1000$    \\
     \hline
     \texttt{$\xi$} & $5$  & $0.5$   & 135 & $1.8\times10^8$  & $1$ & $0.5$    & 50        & 2.05            & 0.1  & $1000$  \\
       & $5$  & $0.5$   & 135 & $3.6\times10^7$  & $5$ & $0.5$    & 50        & 2.05            & 0.1  & $1000$    \\
       & $5$  & $0.5$   & 135 & $1.8\times10^7$  & $10$ & $0.5$    & 50        & 2.05            & 0.1  & $1000$    \\
       & $5$  & $0.5$   & 135 & $0.5\times10^7$  & $36$ & $0.5$    & 50        & 2.05            & 0.1  & $1000$    \\
     \hline
       \texttt{MA} & $5$  & $0.5$   & 135 & $1.5\times10^7$  & $10$ & $0.3$    & 50        & 2.05            & 0.1  & $1000$ \\
       & $5$  & $0.5$   & 135 & $1.5\times10^7$  & $10$ & $0.5$    & 50        & 2.05            & 0.1  & $1000$    \\
       & $5$  & $0.5$   & 135 & $1.5\times10^7$  & $10$ & $0.6$    & 50        & 2.05            & 0.1  & $1000$    \\
     \hline
       \texttt{L0} & $5$  & $0.5$   & 135 & $1.5\times10^7$  & $10$ & $0.5$    & 30        & 2.05            & 0.1  & $1000$\\
       & $5$  & $0.5$   & 135 & $1.5\times10^7$  & $10$ & $0.5$    & 40        & 2.05            & 0.1  & $1000$    \\
       & $5$  & $0.5$   & 135 & $1.5\times10^7$  & $10$ & $0.5$    & 50        & 2.05            & 0.1  & $1000$    \\
       & $5$  & $0.5$   & 135 & $1.5\times10^7$  & $10$ & $0.5$    & 60        & 2.05            & 0.1  & $1000$    \\
     \hline
       \texttt{pe} & $5$  & $0.5$   & 135 & $1.5\times10^7$  & $10$ & $0.5$    & 50        & 1.85            & 0.1  & $1000$\\
       & $5$  & $0.5$   & 135 & $1.5\times10^7$  & $10$ & $0.5$    & 50        & 1.95            & 0.1  & $1000$    \\
       & $5$  & $0.5$   & 135 & $1.5\times10^7$  & $10$ & $0.5$    & 50        & 2.05            & 0.1  & $1000$    \\
       & $5$  & $0.5$   & 135 & $1.5\times10^7$  & $10$ & $0.5$    & 50        & 2.10            & 0.1  & $1000$    \\

    \hline\hline
  \end{tabular}
  }
 \end{center}
 \begin{minipage}{16cm}
 {
 NOTE: All the length scales are in units of $r_{\rm g}={GM_{\bullet}}/{c^2}$, $B_{0}$ in units of Gauss, $n_{\rm e}$ in units of ${\rm cm^{-3}}$, $L_{0}$ the length of the flux rope, $p_{\rm e}$ the index of power-law distributed electrons,  $\eta$ the fraction of thermal energy converted into a population of power-law electrons, $t_{\rm esc}$ in units of minute. \\
 }
 \end{minipage}
\end{table*}


\subsection{Parameter Space Exploration}

In this subsection, we investigate the effects of the individual model parameters given in Table~\ref{tab:parameters} on the resultant light curves. The general approach is that we only modify one parameter a time (except for the concentration parameter $\xi$) from the above fiducial model in order to explore the effects. Below we explore the effect of the following parameters in turn: $B_0$, $n_{\rm e}$, $\xi$, $M_{\rm A}$, $L_0$, and $p_{\rm e}$.

\begin{figure*}
\begin{center}
\includegraphics[width=0.45\textwidth]{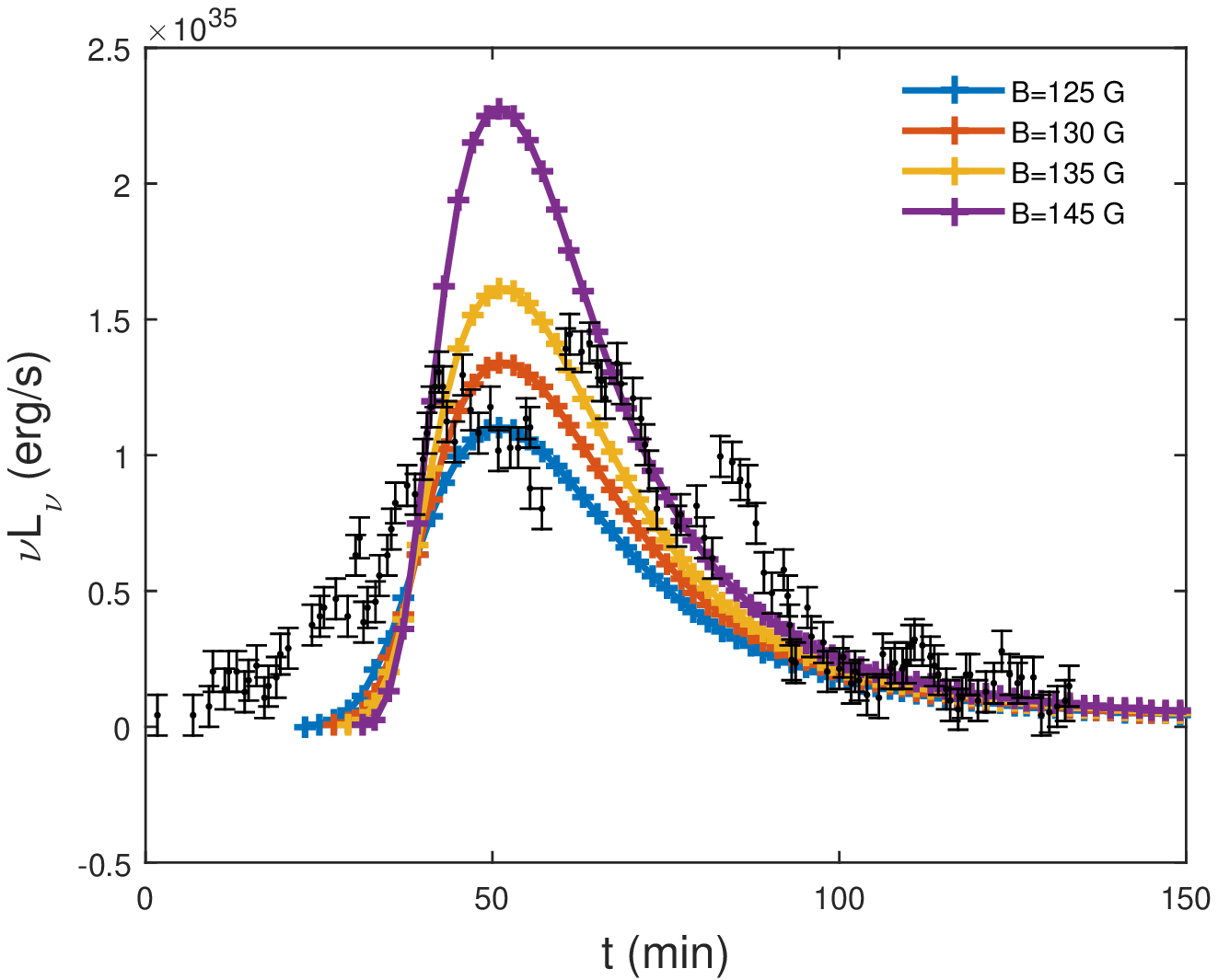}
\includegraphics[width=0.45\textwidth]{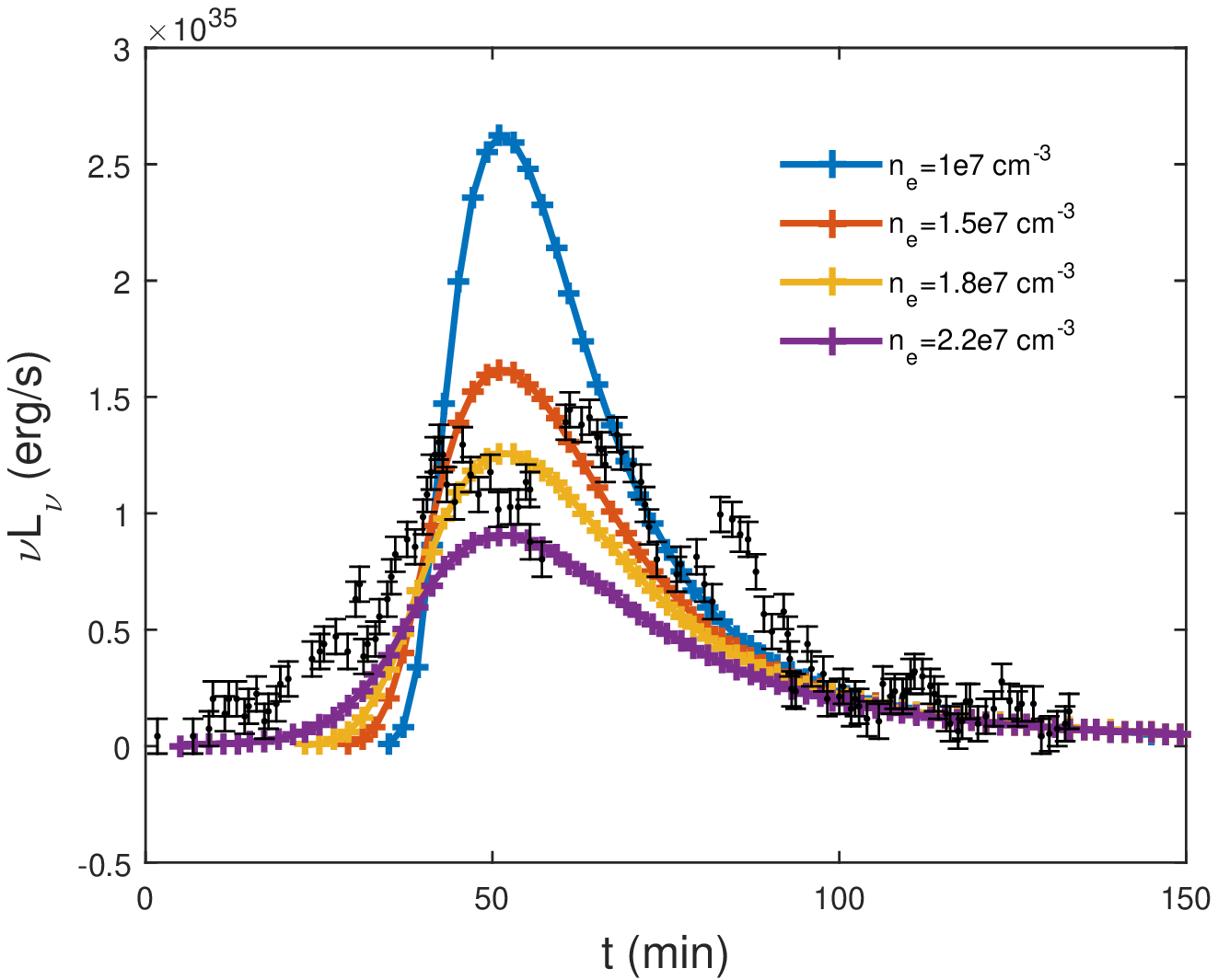}
\includegraphics[width=0.45\textwidth]{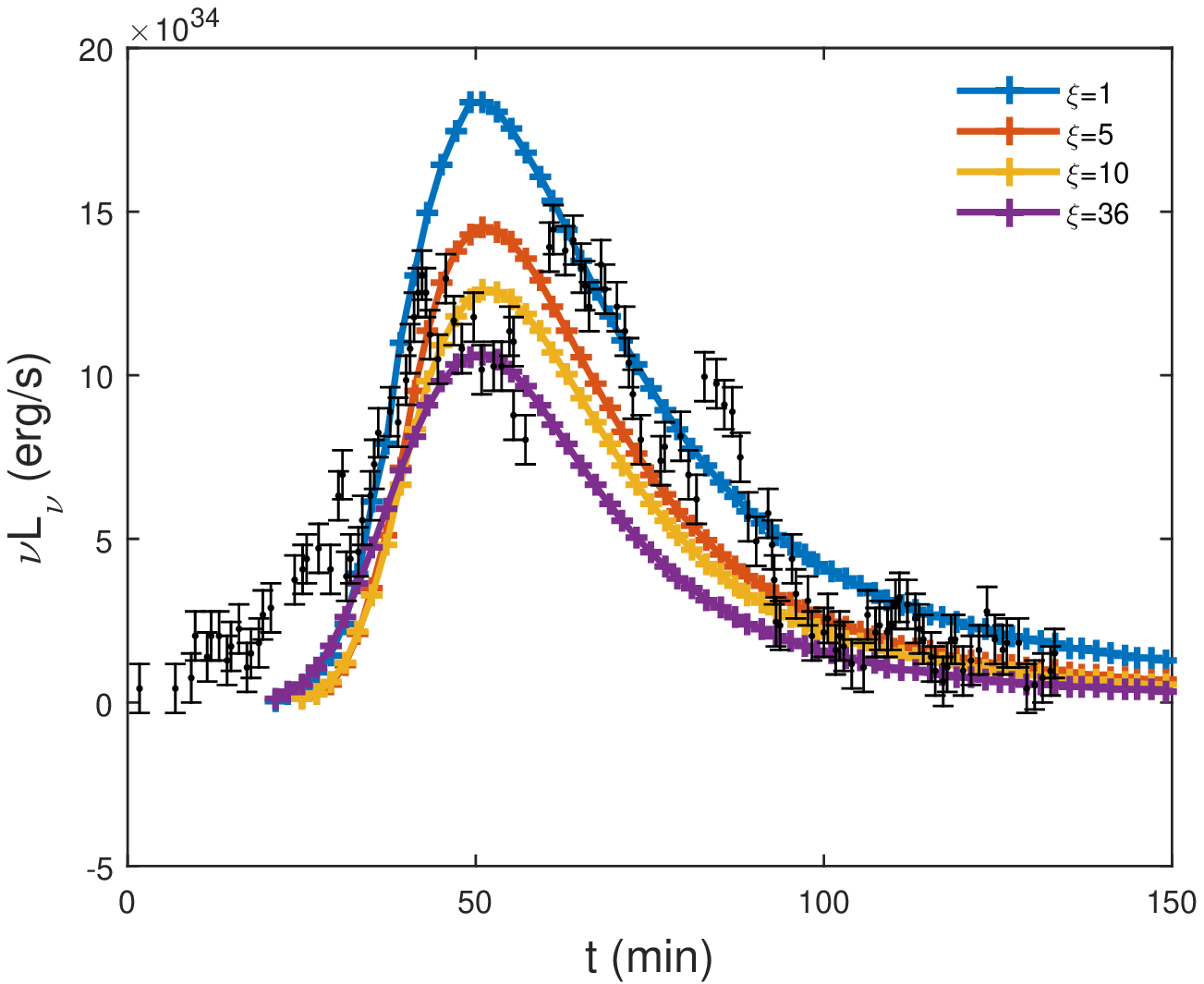}
\includegraphics[width=0.45\textwidth]{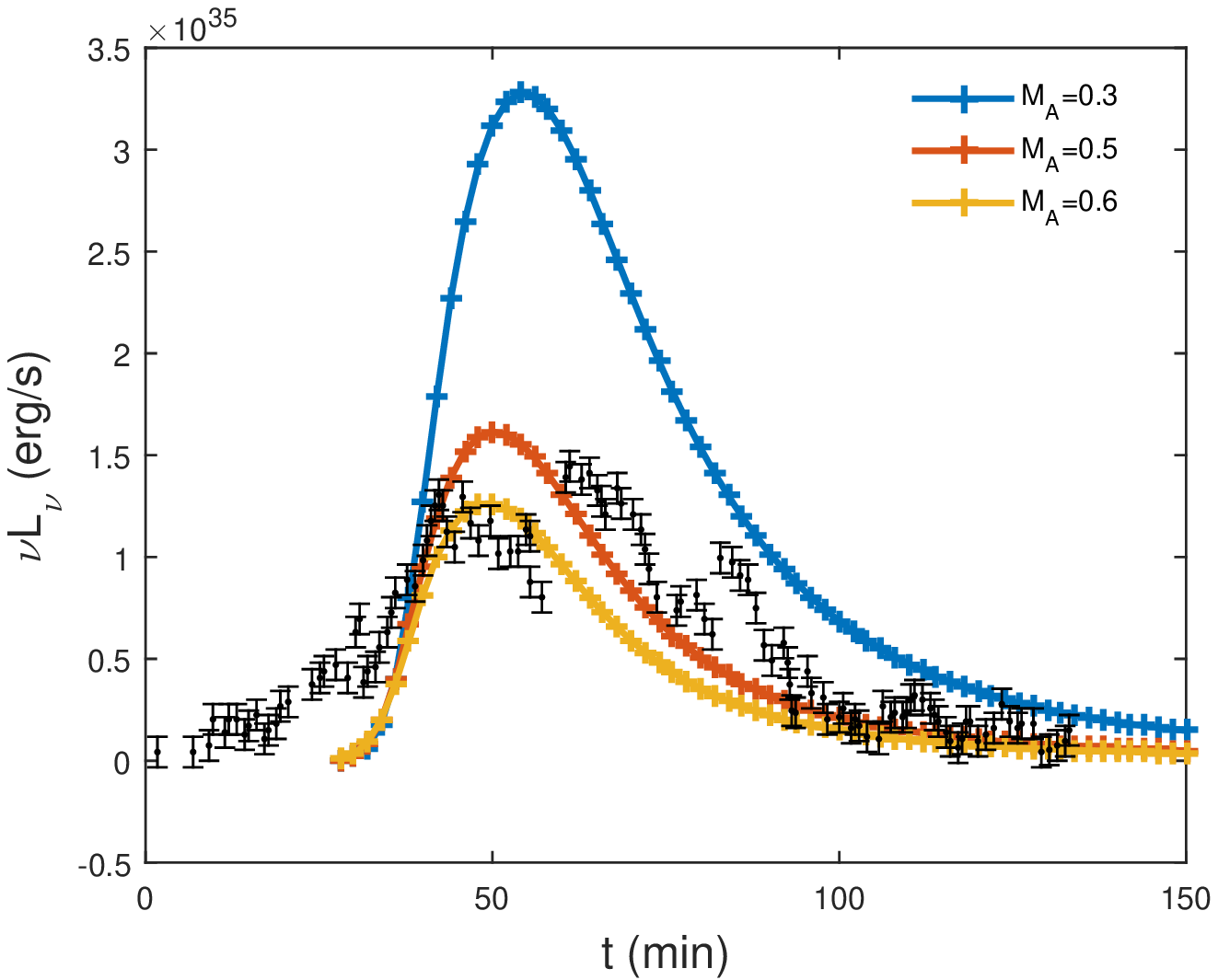}
\includegraphics[width=0.45\textwidth]{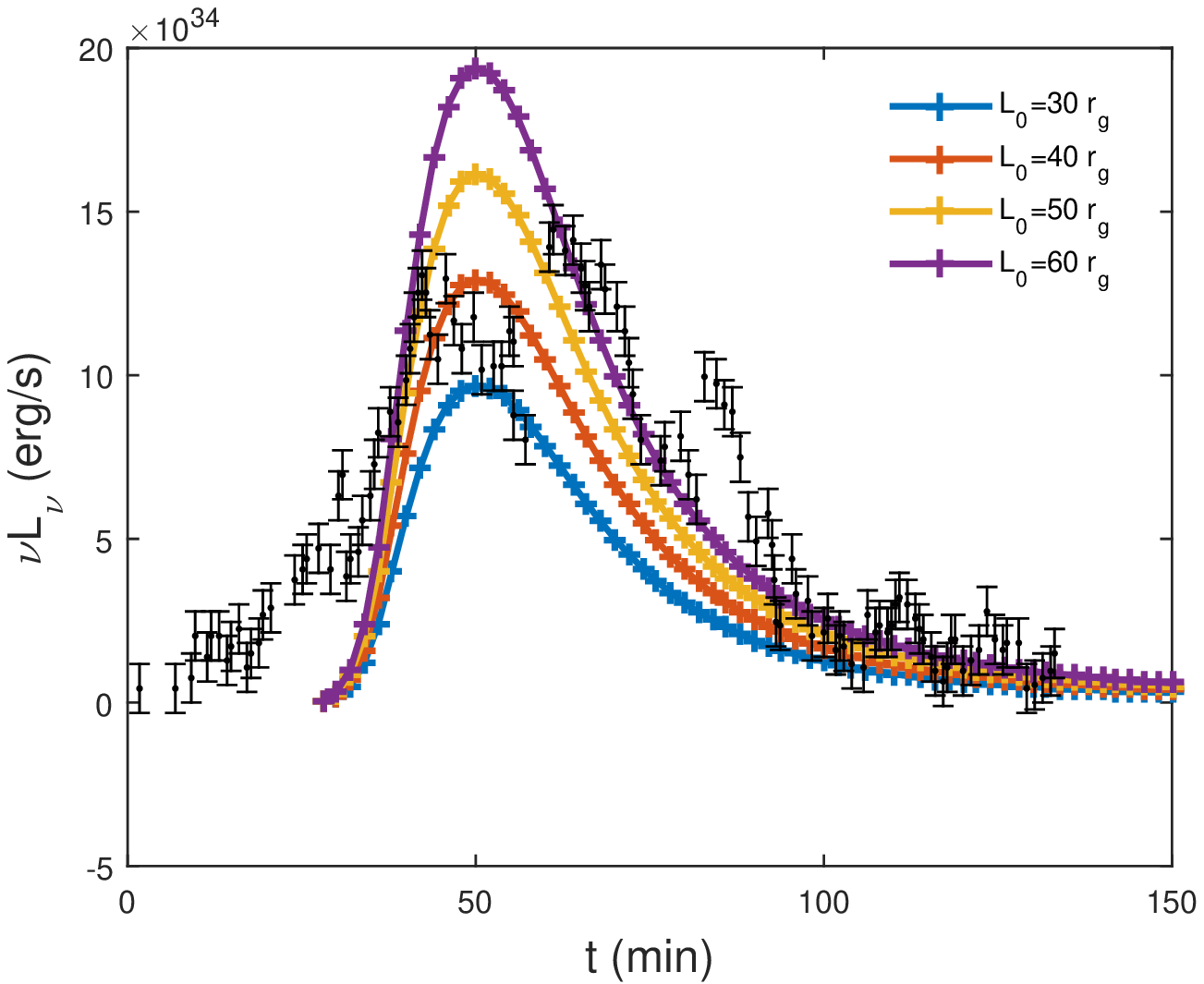}
\includegraphics[width=0.45\textwidth]{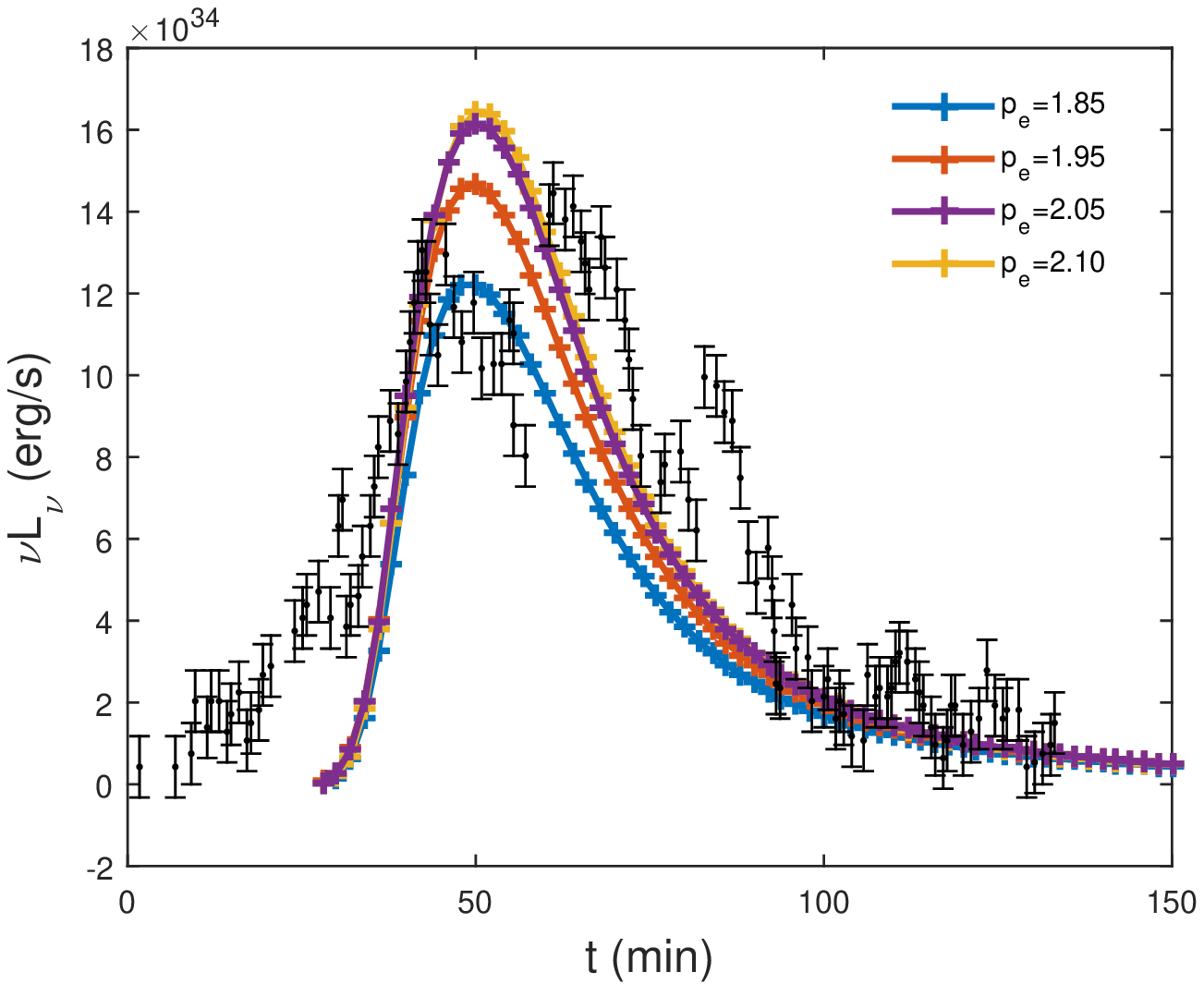}
\end{center}
\caption{The effect of model parameters on the NIR flare light curves. The observed data shown as black points with error bars are superimposed for comparison. For each plot, we only modify one parameter with others fixed. The value for all parameters are listed in Table~\ref{tab:parameters}.}
\label{fig:IRLC}
\end{figure*}

\begin{figure*}
\begin{center}
\includegraphics[width=0.45\textwidth]{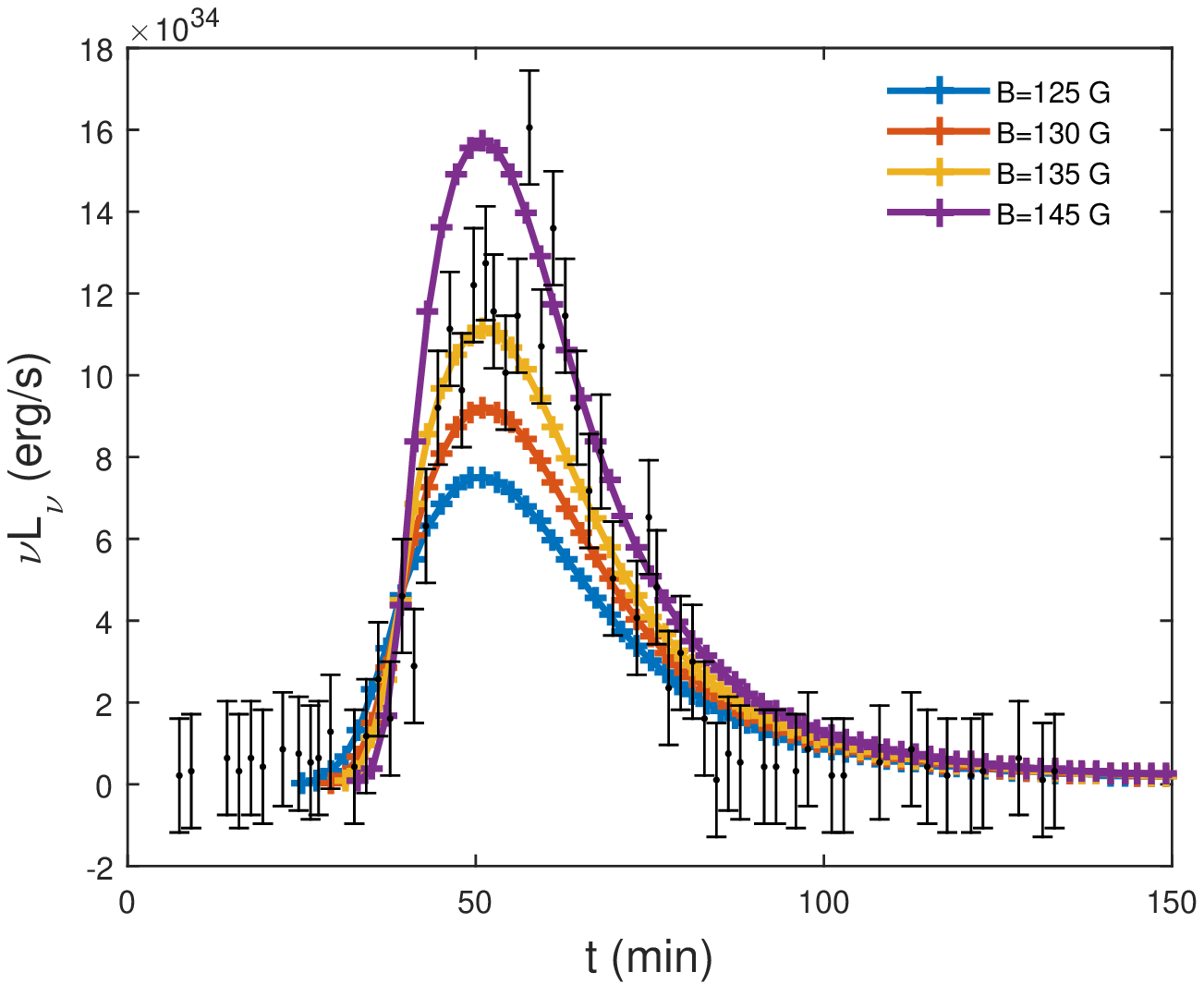}
\includegraphics[width=0.45\textwidth]{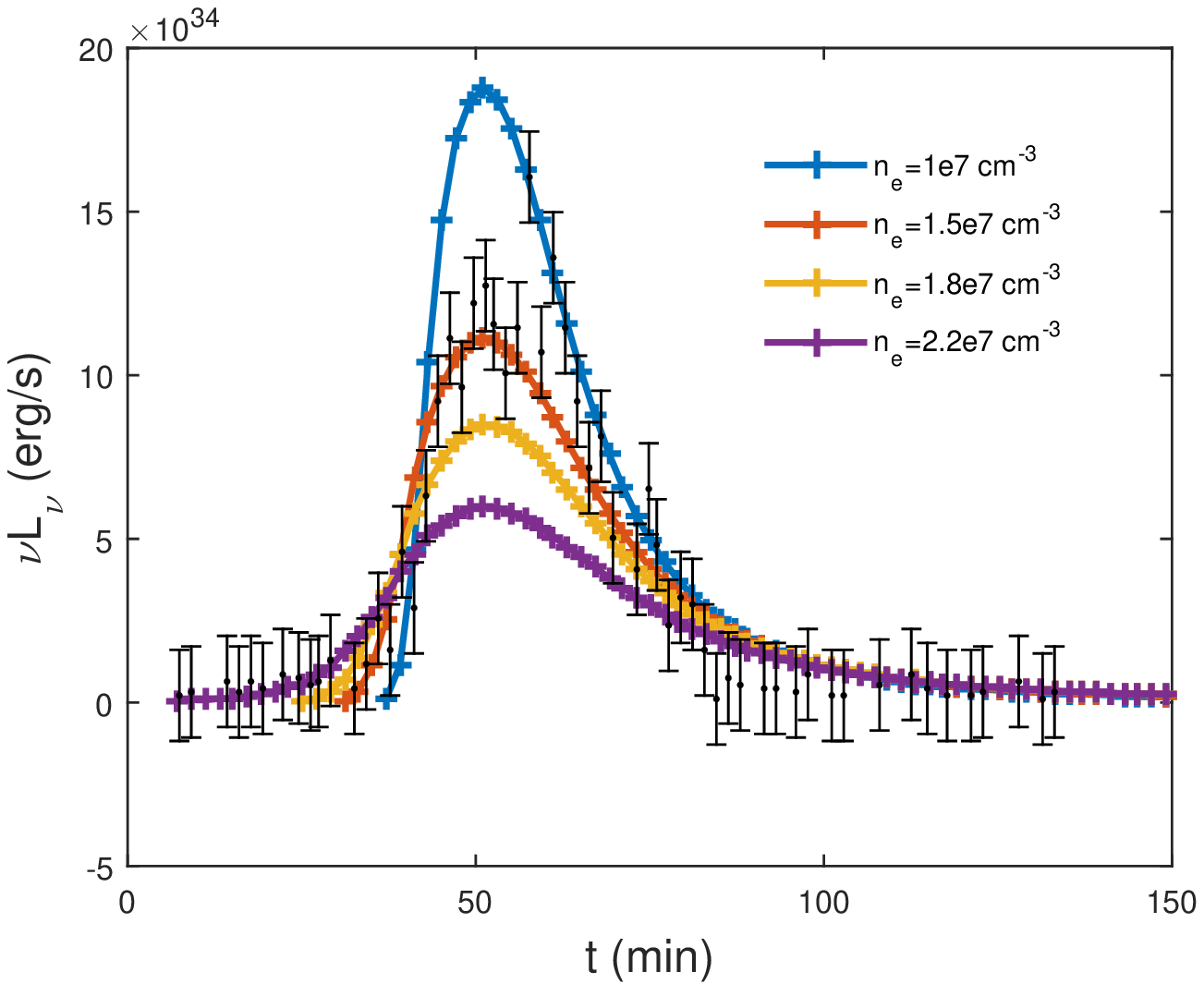}
\includegraphics[width=0.45\textwidth]{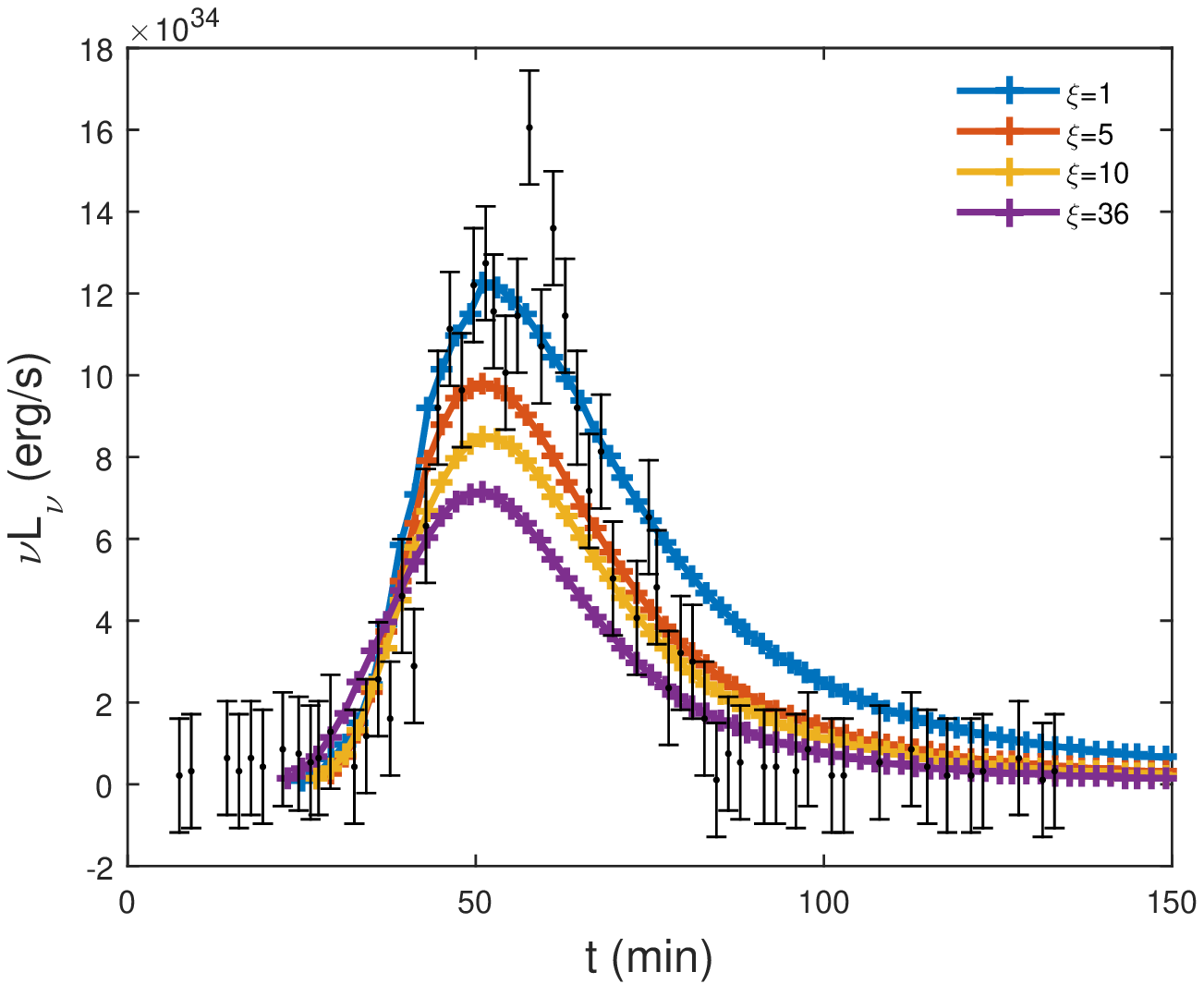}
\includegraphics[width=0.45\textwidth]{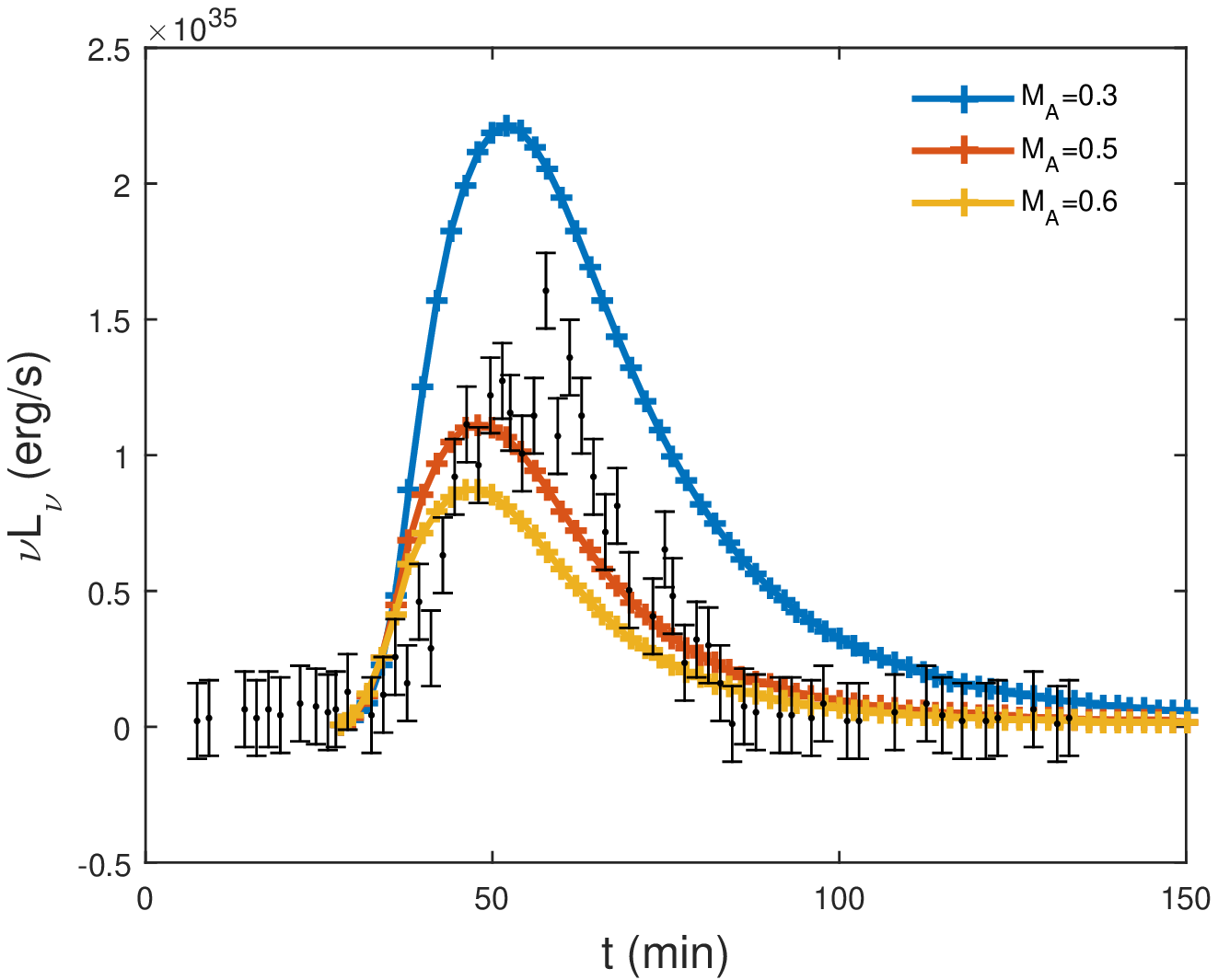}
\includegraphics[width=0.45\textwidth]{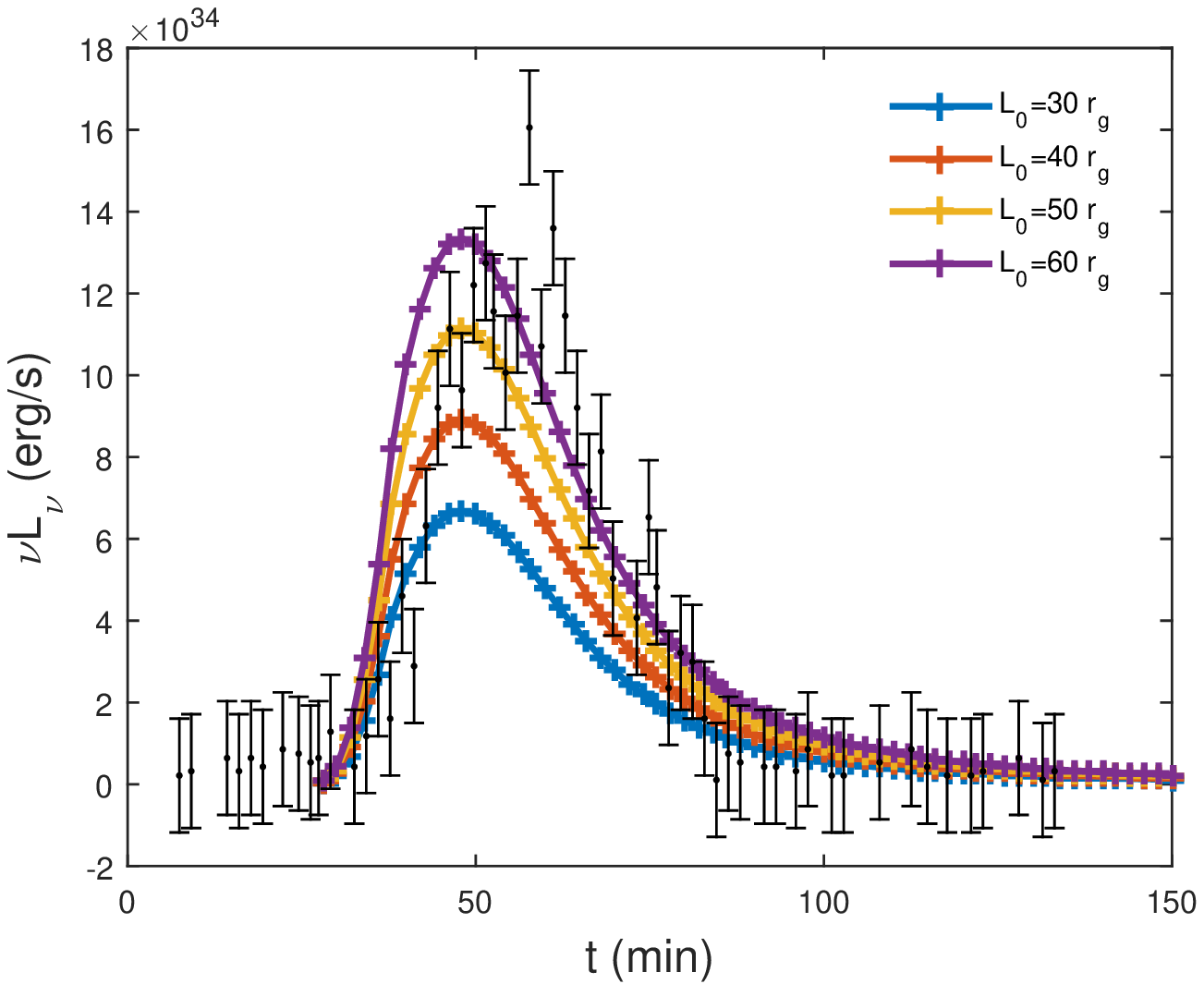}
\includegraphics[width=0.45\textwidth]{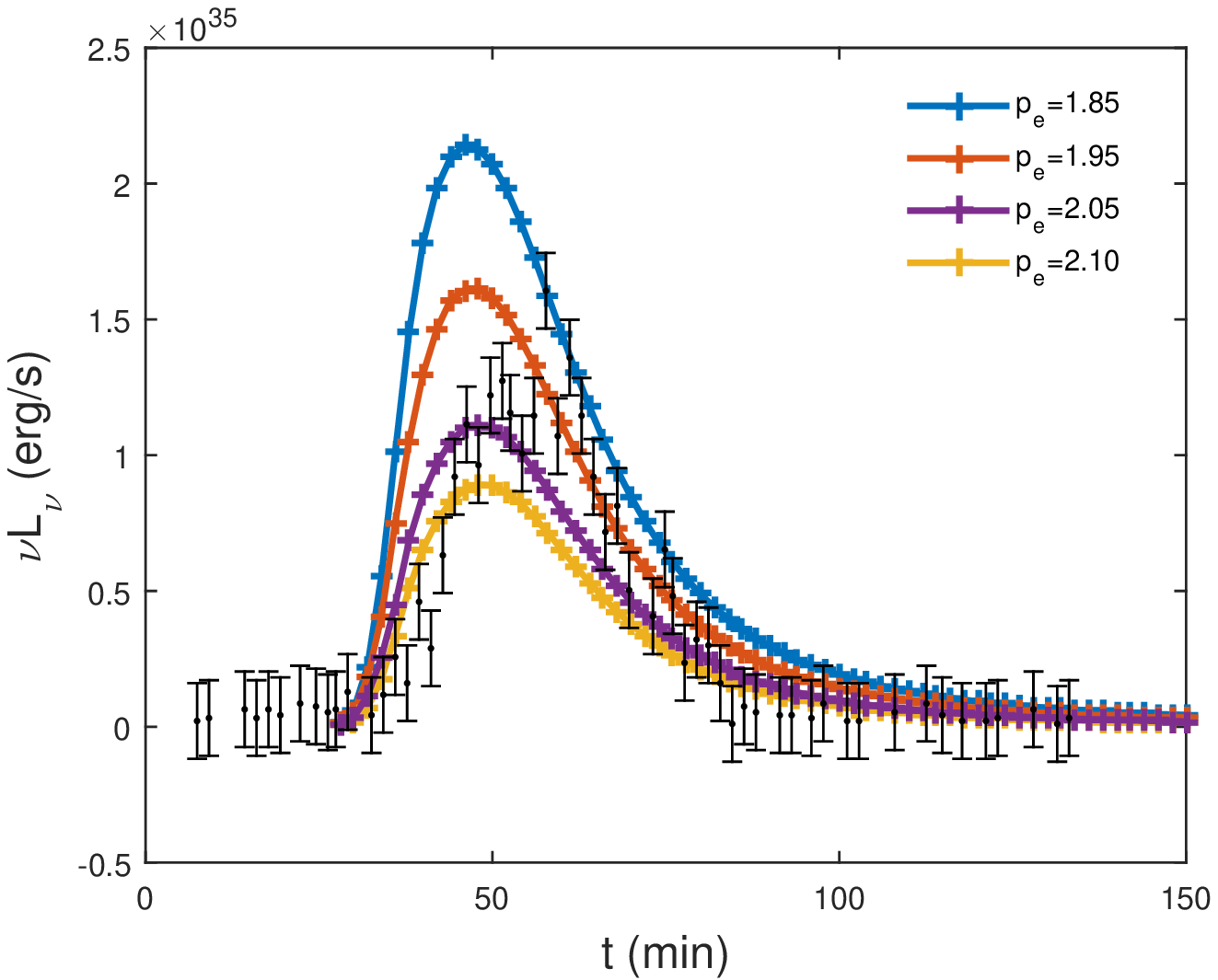}
\end{center}
\caption{Same as Figure~\ref{fig:IRLC} but for X-ray flare light curves. }\label{fig:XLC}
\end{figure*}

In Figure~\ref{fig:IRLC}, we show the results for the effects of the six model parameters on the NIR light curves. It is straightforward to show that the flare luminosity increases with the increasing of the magnetic field strength, $B_{0}$. To compare the flare durations for different $B_{0}$ choices, we align their corresponding light curves with their peak times. As the Alfv\'{e}n timescale $t_{\rm A}\propto{\sqrt{n_{\rm e}}/B_{0}}$, the rise and decline timescales decrease with increasing $B_{0}$, which is due to the shorter Alfv\'{e}n timescale for the stronger magnetic field. One may naturally expect that the duration of a flare would increase with $n_{\rm e}$ as shown in the upper right panel of Figure~\ref{fig:IRLC}. The figure indicates that the flare luminosity is, however, anti-correlated with $n_{\rm e}$. This is  because a lighter blob ejected to farther away from the accretion disc leads to a larger volume of the reconnection region. To investigate the effect of the concentration parameter $\xi$, we make $n_{\rm e}\xi$ a constant to fix the mass of the flux rope. We find that the NIR light curve is insensitive to $\xi$. As shown in the middle left panel of Figure~\ref{fig:IRLC}, the NIR luminosity only slightly decreases with increasing $\xi$. This is because an increase of $\xi$ (or decrease of $n_{\rm e}$) leads to a decrease of the normalization of power-law electrons while the dynamical properties of the flux rope remains the same for a fixed $n_{\rm e}\xi$. As we have stated previously, the parameter $M_{\rm A}$ mainly affects the asymmetry magnitude of the light curve. A larger $M_{\rm A}$ tends to result in a quasi-symmetric profile as shown in the middle right panel of Figure~\ref{fig:IRLC}. The effect of $L_0$ is quite straightforward to understand, i.e., changing the amplitude of the light curve as a whole, as shown in the bottom left panel of Figure~\ref{fig:IRLC}. In the bottom right panel of Figure~\ref{fig:IRLC}, we show the impact of the power-law index $p_{\rm e}$. It is obvious that a harder electron spectrum leads to a weaker NIR flare as well as a stronger X-ray one, as illustrated in the bottom right panel of Figure~\ref{fig:XLC}.

The numerical parameter dependencies  of the X-ray light curves are demonstrated in Figure~\ref{fig:XLC}. For most of the parameters, the quantitative tendency of the X-ray flare luminosity is similar to the case of NIR, but with a weaker dependency. This is due to the fact that the X-ray light curves is mainly determined by the injection profile $c_{\rm inj}$ and is not particularly sensitive to the magnetic field or the minimum Lorentz factor of power-law electrons.

The typical flare light curves that we choose here have comparable amplitudes in NIR and X-ray. As we have reviewed in the Introduction, not every NIR flare has an X-ray counterpart. This observational fact can be explained by either a steeper electron power-law index (as shown in the lower right panel of Figures~\ref{fig:IRLC} and \ref{fig:XLC}) or a smaller maximum Lorentz factor of electrons \citep{Yuan04}. The former could result in a very weak thus undetectable X-ray flare, while in the latter case,  the synchrotron emission does not effectively extend to the X-ray band during NIR flaring activities in observation.

\section{Discussions}

\subsection{Substructures}

An intriguing feature for our selected observations of the April 4 flare is the substructures in the NIR light curve, which is not presented in the X-ray flare observed simultaneously. Although some works suggest that the substructures in NIR flares could be just statistical fluctuations in a red noise spectrum rather than intrinsic flare events \citep{Meyer08,Do09}, it is still interesting to explore the plausible physical mechanism responsible for them.
\citet{DoddsEden10} suggest that this puzzling property is because of the different cooling rate responses of the NIR and X-ray synchrotron-emitting electrons to the change of the magnetic field. In the synchrotron radiation model, small magnetic field fluctuations can produce the substructures observed in the NIR. In contrast, a comparatively smooth shape may be expected in X-ray, because the X-ray emission depends primarily on the injection rate but not on the magnetic field.


During the magnetic reconnection, multiple plasmoids can be formed in the current sheet due to tearing of the current-sheet structure, as seen in numerical simulations \citep{Samtaney09}. When these plasmoids are ejected out from the current sheet, the resultant emission will have substructures in the light curves. This picture is also supported by two-dimensional particle-in-cell simulations of magnetic reconnection \citep{Cerutti12}. These simulations show a strong anisotropy of the particles accelerated by magnetic reconnection and energetic electrons are concentrated into several compact regions inside magnetic islands. The resultant synchrotron light curve of a flare comprises several bright sub-flares emitted by energetic beams of particles. The concentration of the accelerated electrons, which could take in action here in a similar environment, can thus be an alternative mechanism responsible for the substructures observed in the NIR.  The undetected substructures in the X-ray flare at the same level could be explained by the fact that not every ejector can equally contribute to the NIR and X-ray, as we discuss in the Section~\ref{sec:result}. The substructures are also observed in the light curves of the X-ray flares \citep{Baganoff01,Barriere14}. This indicates variability in the particle injection profile. The resultant X-ray light curve may mimic the combined emission from the multiple ejectors.

\subsection{Time Delays}

\citet{Yusef06b} observed that the peak flare emission at 43 GHz leads the 22 GHz peak flare by $\sim20-40$ mins. They  show that the time delay of the flare emission can be naturally interpreted in terms of the plasmon model of \citet{vanLaan66}  by considering the ejection and adiabatic expansion of a uniform, spherical plasma blob.
This is fully consistent with our model, in which flare activities in \sgra\ are associated with blob ejections. In \citet{Yusef06b} the blob ejection is an assumption. In the present work, we provide a dynamical interpretation for such an ejection. The quantitative calculations based on our dynamical model will be presented in a subsequent work.

For the NIR and X-ray considered in this work, the simultaneity is because the flares emissions at these two bands are always optically thin during the whole evolution of the blob. The resultant emissions would thus not suffer from the synchrotron self-absorption effect, which is the reason for the frequency-dependent time lag observed in radio bands \citep{Yusef06b}.



\subsection{Asymmetry}

Although our numerical modellings are focused on the typical light curves which have symmetric profile, there are many events showing asymmetric features. As we have denoted previously, $M_{\rm A}$ affects the magnitude of the profile asymmetry: The observed fast rise and slow decline light curves can be interpreted with $M_{\rm A}<1$, while $M_{\rm A}>1$ tends to generate slow rise and fast decline profiles, as seen in the brightest flare in the XVP campaign \citep{Nowak12}. When $M_{\rm A}$ is about unity, one may expect a rather symmetric light curve, which may then reduce the discrepancy between the rise phase of the light curves as shown in Figure~\ref{fig:lc}. In addition, here we have not taken into account the general relativistical effect. Due to gravitational lensing and time dilution effect, \citet{Younsi15} found that the emitted profile from a ejected plasmoid could be stretched and/or compressed when the plasmoid is close to the central black hole. As a result, asymmetric (fast-rise slow-decay) light curves could lose their original characteristics, and they may even appear to be quasi-symmetric, or as slow-rise fast-decay, as shown in some observations.

\subsection{Polarizations}

The high degree of polarization in NIR is a consequence of the presence of a relatively ordered magnetic field enclosing the ejecta (see Figure~\ref{fig:diagram}) and the small optical depth in the substantially inflated plasmoid blobs. The quantitative discussion of this part will again be presented in the subsequent work. A high degree of polarization for X-ray flares should also be expected due to the same nature as the NIR flares in our model. Such a scenario can thus be tested by the near future X-ray polarimetry, such as \emph{enhanced X-ray Timing and Polarimetry Mission} (\emph{eXTP}).

\section{Summary}

We have developed an analytical MHD model for \sgra\ flares. This work is a development of the \citet{Yuan09} model, which proposed a general scenario for both the formation of episodic ejection of plasmoids from the accretion flow and the associated radiative flares. The model is analogous with the catastrophe model of solar flares and CMEs \citep{LF00}. Theoretically, the analogy between black hole flares and solar flares is based on the similarity of the structure between the accretion flow and the solar atmosphere \citep{Yuan09}. The similarity is further supported by the recent statistical study of X-ray flares, which indicates that they are in a self-organized criticality state driven by magnetic reconnection occurred in the surface of the accretion flow \citep{Li15}.

The basic scenario is briefly summarized as follows. The starting point is a flux rope located in the corona of the accretion flow.  The flux rope is anchored to the accretion flow by the magnetic field lines and is in an equilibrium state initially, balanced by gravity, magnetic tension and pressure forces. The magnetic field lines are controlled by the motion of the accretion flow which is differentially rotating and turbulent. Therefore, magnetic energy and helicity are gradually accumulated with time in the system and eventually reach a threshold. Then the equilibrium of the flux rope is broken down and the flux rope is thrust outwards rapidly. Consequently, the magnetic field lines with opposite directions below the flux rope come close enough, leading to reconnection. Magnetic energy is released in this process and converted into the energy of thermal and power-law electrons. These energetic electrons flow into the flux rope and the magnetic loops where they then emit strong synchrotron radiation. This can explain the observed flares. In this scenario, the radiative flares are associated with expanding hot spots close to the black hole, which could be tested by future observations. The schematic figure of the process is shown in Figure~\ref{fig:diagram}.

By assuming certain spacial distributions of the magnetic field and density in the coronal region, we have calculated the dynamical evolution of the height of flux rope (Figure~\ref{fig:dyn}), the Alfv\'{e}n speed (Figure~\ref{fig:VA}), the expansion of the flux rope (Figure~\ref{fig:ad}), the magnetic field close to the reconnection region (Figure~\ref{fig:magB}), the released energy in the reconnection (Figure~\ref{fig:Edot}), and the minimum Lorentz factor of accelerated electrons (Figure~\ref{fig:cinj}).  The dynamical evolution of these parameters further allow us to calculate the evolution of the energy distribution of accelerated electrons in the current sheet (Figure~\ref{fig:Ne}), and further their radiation. The results of these calculations are then compared with light curves and SED observed on April 4, 2007 (Figures~\ref{fig:lc} \& \ref{fig:sed}). Our numerical results show that the flux rope ejected from the surface of the accretion flow can be accelerated to mildly relativistic velocity within $\sim1\ {\rm hr}$ after the loss of equilibrium. With the relatively large Alfv\'{e}n Mach number $M_{\rm A}$, the reconnection of the current sheet can be very efficient and thus the bottom and top tips of the current sheet are rather close to each other. (Figure~{\ref{fig:dyn}}). As the reconnection proceeds, large amounts of the energy flux and particles are brought into the current sheet, and then ejected upwards and downwards. One half of the Poynting flux and energetic electrons are flowing into the flux rope and the flare loop, where flaring activities take place. With appropriate choices of the parameters, we find that the total power by the magnetic reconnection can reach $\sim10^{37}\ {\rm erg\ s^{-1}}$, and the particle injection rate into the flare regions $\sim10^{43}\ {\rm s^{-1}}$ (Figure~\ref{fig:Edot}). The radiative efficiency is thus only $\sim1\%$ for the resultant luminosity of $10^{35}\ {\rm erg\ s^{-1}}$.

Our calculation results can reasonably explain the main characteristics of the observed flares, including their IR and X-ray light curves (Figure~\ref{fig:lc}) and the spectra (Figure~\ref{fig:sed}). The model can explain not only why NIR and X-ray flares occur simultaneously if both of them are observed, but also some of the NIR flares do not have corresponding X-ray counterparts. Moreover, the astrometric signatures during strong flares due to the expansion and/or ejection blob near the black hole could be detected by future high spatial resolution instruments, such as VLTI GRAVITY \citep{Eisenhauer11}.
The scenario of an expanding radio-emitting blob can also naturally explain the observed time lag between the two light curves at two radio frequencies \citep[e.g.,][]{Yusef06b,Brinkerink15}.
The quantitative calculation, together with the interpretation of the observed NIR polarization will be presented in a subsequent paper. Our model also predicates a high degree of polarization for X-ray flares, which can be verified by the near future X-ray polarimetry, e.g., \emph{eXTP}.

\section*{Acknowledgments}
We are deeply grateful to Jun Lin and Ying Meng for help in understanding solar flare/CME models. We would like to thank an anonymous referee for the useful suggestions to improve the manuscript. This work is supported in part by the Natural Science Foundation of China (grants 11573051 and 11633006), the Key Research Program of Frontier Sciences of CAS (No.  QYZDJ-SSW-SYS008), the grant from the Ministry of Science and Technology of China (No. 2016YFA0400704), and the CAS/SAFEA International Partnership Program for Creative Research Teams. Y.P.L. is also sponsored in part by Shanghai Sailing Program (No. 17YF1422600).

\appendix
\section{Solving Equation~(1)}

The other terms involved in Equations~(\ref{eq:dpdh},\ref{eq:dqdh},\ref{eq:A0hp}) are listed as follows:
\begin{eqnarray}\label{eq:A0}
{A_{\rm R}} &= &\frac{{\lambda {H_{\rm PQ}}}}{{2h{L_{\rm PQ}}}}
\ln \left[ \frac{{\lambda {H^3_{\rm PQ}}}}{{{r_{00}}{L_{\rm PQ}}({h^4} - {p^2}{q^2})}} \right]  \nonumber \\
  &+& \tan^{-1}\left( \frac{\lambda }{h}\sqrt {\frac{{{p^2} + {\lambda ^2}}}{{{q^2} + {\lambda ^2}}}} \sqrt {\frac{{{h^2} - {q^2}}}{{{h^2} - {p^2}}}} \right) \nonumber \\
  &+& \frac{\lambda }{{q{L_{\rm PQ}}}} \left\{ ({h^2} - {q^2})F\left[ {\sin^{-1}}\left(\frac{q}{h}\right),\frac{p}{q} \right] \right. \nonumber \\
  &+& ({q^2} - {p^2})\Pi \left[ {\sin^{-1}}\left(\frac{q}{h}\right),\frac{{{p^2} + {\lambda ^2}}}{{{q^2} + {\lambda ^2}}},\frac{p}{q} \right] \nonumber \\
  &-& \left. \frac{{{H^2_{\rm PQ}}}}{{{h^2}}}\Pi \left[ {\sin^{-1}}\left(\frac{q}{h}\right),\frac{{{p^2}}}{{{h^2}}},\frac{p}{q} \right]  \right\} \nonumber \\
  &=& \frac{\pi }{4} + \ln \left( \frac{{2\lambda }}{{{r_{00}}}} \right), \nonumber \\
{A_{\rm Rp}} &= & \frac{{\lambda p({h^2} + {\lambda ^2})}}{{q{{({p^2} + {\lambda ^2})}^2}}}\sqrt {\frac{{{p^2} + {\lambda ^2}}}{{{q^2} + {\lambda ^2}}}} \left\langle \left(1 - \frac{{{q^2}}}{{{h^2}}} \right) \right. \nonumber \\
 &\times& \Pi \left[{\sin^{-1}}\left(\frac{q}{h}\right),\frac{{{p^2}}}{{{h^2}}},\frac{p}{q} \right]  \nonumber \\
  &-& F \left[ {\sin^{-1}}\left(\frac{q}{h}\right),\frac{p}{q} \right] \nonumber \\
  &-& \left. \frac{q}{{2h}}\sqrt {\frac{{{h^2} - {q^2}}}{{{h^2} - {p^2}}}} \left\{1 + \ln \left[ \frac{{\lambda {H^3_{\rm PQ}}}}{{{r_{00}}{L_{\rm PQ}}({h^4} - {p^2}{q^2})}}\right] \right\} \right\rangle, \nonumber \\
 {A_{\rm Rq}} &=& \frac{{\lambda ({h^2} + {\lambda ^2})}}{{{{({q^2} + {\lambda ^2})}^2}}}\sqrt {\frac{{{q^2} + {\lambda ^2}}}{{{p^2} + {\lambda ^2}}}} \left\langle \left(1 - \frac{{{p^2}}}{{{h^2}}} \right) \right. \nonumber \\
 &\times& \Pi \left[ {\sin^{-1}}\left(\frac{q}{h}\right),\frac{{{p^2}}}{{{h^2}}},\frac{p}{q} \right] \nonumber\\
 &-&  F\left[ {\sin^{-1}}\left(\frac{q}{h}\right),\frac{p}{q} \right] \nonumber \\
 &-& \left. \frac{q}{{2h}}\sqrt {\frac{{{h^2} - {p^2}}}{{{h^2} - {q^2}}}} \left\{1 + \ln \left[\frac{{\lambda {H^3_{\rm PQ}}}}{{{r_{00}}{L_{\rm PQ}}({h^4} - {p^2}{q^2})}}\right] \right\} \right\rangle, \nonumber \\
 \nonumber \\
 {A_{\rm Rh}} &= &\frac{\lambda }{{2{h^2}{L_{\rm PQ}}{H_{\rm PQ}}}}\left\{ 2\frac{{{h^6} - {\lambda ^2}{p^2}{q^2}}}{{{h^2} + {\lambda ^2}}} \right. \nonumber \\
  &-&   \frac{{h^2}({p^2} + {q^2})({h^2} - {\lambda ^2})}{{h^2} + {\lambda ^2}}   \nonumber \\
  &+& ({h^4} - {p^2}{q^2})\left. \ln\left[ \frac{{\lambda {H^3_{\rm PQ}}}}{{{r_{00}}{L_{\rm PQ}}({h^4} - {p^2}{q^2})}} \right] \right\}  \nonumber \\
  &+&  \frac{\lambda }{{hq{L_{\rm PQ}}}}\left\{ ({h^2} + {q^2})F\left[{\sin^{-1}}\left(\frac{q}{h}\right),\frac{p}{q}\right] \right. \nonumber \\
  &-& {q^2}E\left[{\sin^{-1}}\left(\frac{q}{h}\right),\frac{p}{q}\right] \nonumber \\
  &-& \left. \frac{{{h^4} - {p^2}{q^2}}}{{{h^2}}}\Pi \left[ {\sin^{-1}}\left(\frac{q}{h}\right),\frac{{{p^2}}}{{{h^2}}},\frac{p}{q}\right] \right\}.
\end{eqnarray}
and
\begin{eqnarray}\label{eq:AR}
 A_{0}^{0} &=& \frac{{2{I_0}}}{c}\frac{\lambda }{{q{L_{\rm PQ}}}} \left[ ({h^2} - {q^2})K\left(\frac{p}{q}\right) + ({q^2} - {p^2}) \right. \nonumber \\
  &\times & \Pi \left(\frac{{{p^2} + {\lambda ^2}}}{{{q^2} + {\lambda ^2}}},\frac{p}{q}\right) - \left. \frac{{{H^2_{\rm PQ}}}}{{{h^2}}}\Pi \left(\frac{{{p^2}}}{{{h^2}}},\frac{p}{q} \right) \right], \nonumber\\
 {A_{\rm 0p}}& = & \frac{{\lambda p({h^2} + {\lambda ^2})({q^2} + {\lambda ^2})}}{{{h^2}q{{[({p^2} + {\lambda ^2})({q^2} + {\lambda ^2})]}^{3/2}}}} \nonumber \\
  & \times & \left[ ({h^2} - {p^2})\Pi \left(\frac{{{p^2}}}{{{h^2}}},\frac{p}{q}\right) -
  {h^2}K\left(\frac{p}{q}\right) \right],\nonumber \\
 {A_{\rm 0q}}&= &\frac{{\lambda ({h^2} + {\lambda ^2})({p^2} + {\lambda ^2})}}{{{h^2}{{[({p^2} + {\lambda ^2})({q^2} + {\lambda ^2})]}^{3/2}}}} \nonumber \\
  &\times & \left[ ({h^2} - {q^2})\Pi \left(\frac{{{p^2}}}{{{h^2}}},\frac{p}{q}\right) -
  {h^2}K\left(\frac{p}{q}\right) \right], \nonumber\\
 {A_{\rm 0h}} &= & - \frac{\lambda }{{{h^3}q\sqrt {({p^2} + {\lambda ^2})({q^2} + {\lambda ^2})} }} \nonumber \\
  &\times & \left[ {h^2}{q^2}E\left(\frac{p}{q}\right) - {h^2}({h^2} + {q^2})K\left(\frac{p}{q}\right) \right. \nonumber \\
 & + & \left. ({h^4} - {p^2}{q^2})\Pi \left(\frac{{{p^2}}}{{{h^2}}},\frac{p}{q}\right) \right].
 \\ \nonumber
\end{eqnarray}
Here $K$, $E$ and $\Pi$ in Equation~(\ref{eq:A0}) are first, second, and third kinds of complete elliptic integrals, respectively; $F$, $E$ and $\Pi$ in Equation~(\ref{eq:AR}) are first, second, and third kinds of incomplete elliptic integrals, respectively.


\bsp	
\label{lastpage}

\end{document}